\documentclass[epj]{svjour}
\usepackage{graphicx}
\usepackage{amssymb,amsbsy,amsmath}
\newcommand{\bs}{\boldsymbol}
\def\be{\begin{equation}}
\def\ee{\end{equation}}
\def\beq{\begin{eqnarray}}
\def\eeq{\end{eqnarray}}
\def\bc{\begin{center}}
\def\ec{\end{center}} 
\newcommand{\nn}{\nonumber\\}
\journalname{Eur. Phys. J. B}
\begin{document}
\title{Phase diagram of the alternating-spin Heisenberg chain with extra isotropic
three-body  exchange interactions}
\titlerunning{Phase diagram of the alternating-spin Heisenberg chain}
\author{Nedko B. Ivanov%
\inst{1,2}
 \and J\"org Ummethum%
\inst{1}
 \and J\"urgen Schnack%
\inst{1}
}                     % Do not remove
\offprints{Nedko B. Ivanov}          % Insert a name or remove this line
\institute{Department of Physics, Bielefeld University, P.O. box
  100131, D-33501 Bielefeld, Germany \and Institute of Solid
  State Physics, Bulgarian Academy of Sciences, Tzarigradsko
  chaussee 72, 1784 Sofia, Bulgaria} 
\date{Received: date / Revised version: date}
\abstract{
For the time being isotropic three-body exchange interactions are 
scarcely  explored and mostly used as a tool for constructing various  
exactly solvable  one-dimensional models, although, generally speaking, 
such competing terms in generic Heisenberg  spin systems  
can be expected to  support specific quantum effects and phases. 
The Heisenberg chain constructed from alternating $S=1$ and
$\sigma=\frac{1}{2}$ site spins defines a realistic prototype model 
admitting extra three-body exchange terms. Based on numerical  density-matrix 
renormalization group (DMRG) and exact diagonalization (ED) calculations, 
we demonstrate that the  additional isotropic three-body terms stabilize    
a variety of  partially-polarized states  as well as
two  specific non-magnetic states including a critical  spin-liquid phase   
controlled by two Gaussinal conformal theories as well as a critical nematic-like phase 
characterized by dominant quadrupolar
$S$-spin fluctuations.  Most of the established  effects are related to some 
specific features  of the  three-body interaction  such as  the promotion 
of local collinear spin configurations and the enhanced tendency towards  
nearest-neighbor clustering of the spins. It may be expected that most of the predicted 
effects of the isotropic three-body interaction 
persist  in higher space dimensions.      
\PACS{
{75.10.Jm}{Quantized spin models}   \and
{75.40.Mg}{Numerical simulation studies}   \and
{75.45.+j}{Macroscopic quantum phenomena in magnetic systems}
     } % end of PACS codes
} %end of abstract
\maketitle
%
%%%%%%%%%%%%%%%%%%%%%%%%%
\section{Introduction }
%%%%%%%%%%%%%%%%%%%%%%%%%%
For the past two decades, it has been demonstrated that the frustrated magnetic systems
host a rich variety of new macroscopic states. In addition to various  geometrically frustrated 
(triangular type) lattices, competing interactions in the Heisenberg spin 
models--such as longer-range bilinear exchange terms, 
the Dzyaloshinskii-Moria interaction, as well as different
ring and biquadratic  exchange 
couplings--have been widely discussed as sources of exotic non-magnetic
quantum states, including different spin-liquid, nematic, and topological
phases \cite{frustration}.  Heisenberg spin models  with  two-site  biquadratic
terms,  $\left(\bs{S}_i\cdot\bs{S}_j\right)^2$, are among the
most-often studied  spin systems   
with higher-order exchange interactions. Typical examples with rich phase diagrams  are  
the spin-1 bilinear-biquadratic (BBQ) chain \cite{spin_1_chain} 
and its higher-dimensional counterparts on square \cite{harada,spin_1_2D},
triangular \cite{momoi,smerald}, and cubic \cite{harada} lattices. The phase diagram of the 
BBQ chain contains two gapped (one Haldane
and one dimerized) states and an exotic critical phase characterized by
nematic  spin-spin correlations with the  dominant momenta  $k=\pm\frac{3\pi}{4}$, 
whereas the 2D square-lattice analogues support a number of exotic nematic phases. 

In contrast to the pronounced  interest in  biquadratic couplings,  
by now the role of the isotropic three-site exchange   
$\left( \bs{S}_i\cdot\bs{S}_j\right)\left(
\bs{S}_i\cdot\bs{S}_k\right)+h.c.$ ($|\bs{S}_i|>\frac{1}{2}$, $i\neq j, k$,
$j\neq k$) remains  scarcely  explored. Although the  two-body interactions  play
a fundamental role, the search of  systems described by 
effective many-body Hamiltonians can be  motivated by the  expected 
specific effects and exotic phases is such systems. In principle, 
it is difficult to identify real physical systems exhibiting properties
related to such models. To the best of our knowledge, the only  more convincing experimental 
evidence for effects related to three-body spin interactions  comes from  inelastic neutron scattering results
for the  low-lying  excitations in the magnetic material CsMn$_x$Mg$_{1-x}$Br$_3$
($x=0.28$) \cite{falk1}, CsMnBr$_3$ being  known as a nearly ideal isotropic 1D Heisenberg
antiferromagnet with site spins  $S=\frac{5}{2}$.  These experimental 
results  predicted  almost identical strengths of both the
biquadratic and three-site interactions,  which are about two orders of 
magnitude weaker than the principal 
Heisenberg coupling. The higher-order  spin-spin interactions in   CsMn$_x$Mg$_{1-x}$Br$_3$
appear  as a result of magenetoelastic forces \cite{falk2}. 
Similar magnetostriction effects -- earlier discussed for  polynuclear
complexes of iron-group ions \cite{iwashita} -- were predicted for 
some single-molecular magnets \cite{furrer}.    
Both types of higher-oder exchange interactions also naturally appear 
in the fourth order of the strong-coupling expansion of the 
two-orbital Hubbard model \cite{michaud1}.        
However, in both models the strengths of these 
interactions are controlled by one and the same model parameter, so
that it might  be  difficult to isolate the  effects related to different
higher-order terms in the Hamiltonian. Therefore, another  challenge in 
the field is  to identify  experimentally accessible 
systems  where the effects of higher-order   
interactions can be definitely isolated. 
Cold atoms in optical lattices open a promising route in this direction. 
It has been demonstrated \cite{pachos} that with the  two-species 
Bose-Hubbard model  in a triangular configuration a wide range of Hamiltonian 
operators could be generated that include effective three-spin interactions. 
The latter result from the possibility of atomic tunneling
through different paths from one vertex to 
another one, and can be extended to  1D spin models with
three-spin 
interactions. Another intriguing system in optical lattices --
opening a route for experimental  studies of the three-body
interactions -- concerns  polar molecules   
driven by microwave fields, naturally giving rise to Hubbard models with
strong nearest-neighbor three-body interactions \cite{buchler}.

For the time being isotropic three-body exchange interactions are
 mostly used as a tool for constructing various  
exactly solvable one-dimensional (1D)
 models \cite{andrej,devega_woynar,aladim,devega,bytsko,ribeiro}.
Only recently some specific features of the three-body 
exchange interaction  in generic spin-S Heisenberg 
models in space dimensions D=1 and 2 have been discussed
in the literature \cite{michaud1,michaud2,michaud3,wang}. In particular, it has
been argued that for some  strengths of this interaction  the spin-S Heisenberg chain  
exhibits  an exact fully-dimerized (Majumdar-Ghosh type)
 ground state (GS) \cite{michaud1,wang}.
The numerical results for $S=1$, $\frac{3}{2}$, and $2$ support the suggestion  that the 
related dimerization transition in this system is described by the 
$SU(2)_{k}$ Wess-Zumino-Witten  model with  the central charge
$c=3k/(2+k)$, where $k=2S$ \cite{michaud2}. In addition, another recent work of these
authors demonstrated  a rich variety of phases in the phase diagram  of the spin-1 
Heisenberg model on a square lattice with extra isotropic 
three-body exchange interactions \cite{michaud3}.      

In the framework of spin systems on conventional lattices, some  systems
described by Heisenberg alternating-spin  models seem to suggest another 
realistic onset for observing and separating the effects of   higher-order exchange
interactions. The Heisenberg chain with alternating $S$ and $\sigma=\frac{1}{2}$ spins
($S>\frac{1}{2}$) provides  a simple  example of this kind. Indeed,
according to the operator identity $\left( \bs{S}_i\cdot\sigma_j\right)^2\equiv 
-\bs{S}_i\cdot\sigma_j/2+S(S+1)/4$, the  biquadratic terms
in this system reduce to  bilinear forms. In view of the numerous experimentally 
accessible quasi-1D  spin systems described by the  Heisenberg model with
alternating spins \cite{furrer,landee},  in this work we concentrate on a
generic 1D model of this class defined by the following  Hamiltonian
\beq\label{h}
{\cal H}_{\sigma S}&=&\sum_{n=1}^Lh_n\equiv  \sum_{n=1}^L J_1 \bs{S}_{2n}\!\cdot\!\left(
\bs{\sigma}_{2n-1}\!+\!\bs{\sigma}_{2n+1}\right)\nn
&+& J_2\left[\left(\bs{S}_{2n}\!\cdot\!\bs{\sigma}_{2n-1}\right)
\left( \bs{S}_{2n}\!\cdot\!\bs{\sigma}_{2n+1}\right)\!+\! h.c.\right].
\eeq
Here  $L$ stands for the number of elementary cells, 
each containing two different spins ($S>\sigma$). We shall use  the 
standard parameterization of the coupling constants 
$J_1=\cos (t)$ and  $J_2=\sin (t)$ ($0\leq t<2\pi$). 
Since the effective strength of
the extra term is controlled by the parameter $S\sigma J_2$, it is reasonable to
expect that  this interaction could play an important role especially
in  $(S,\frac{1}{2})$ chains and rings  with large $S$ spins ($S\gg
\frac{1}{2}$).  In  the extreme quantum case $(S,\sigma)=(1,\frac{1}{2})$,
${\cal H}_{\sigma S}$ reproduces (up to irrelevant constants) 
the  effective Hamiltonian of the   isotropic  spin-$\frac{1}{2}$ diamond chain 
 (with  an additional ring exchange in the plaquettes) in the Hilbert subspace 
where the  pairs  of "up" and "down" plaquette spins 
form triplet states \cite{ivanov1}.

The paper is organized as follows. In Sec.~\ref{sec_cpd} we
discuss the classical phase diagram of 
the model, whereas Section~\ref{sec-exact} contains some exact
analytical results concerning 
the three-site cluster, as well as the one-magnon excited states and
phase boundaries of the  FM phase. In Sec.~\ref{sec_qpd} we present the
quantum phase diagram of the model for the extreme quantum case ($S=1$ and
$\sigma =\frac{1}{2}$), based on numerical DMRG as well as
exact-diagonalization (ED) simulations,  
and discuss different  properties of the phases. The last Section contains a
summary of the results.   If  not specially mentioned, the results in 
the following Sections concern the extreme quantum case 
$S=1$ and $\sigma=\frac{1}{2}$. 

%%%%%%%%%%%%%%%%%%%%%%%%%%%%%%%%%%%%%%%%%%%%%%%%%%
\section{\label{sec_cpd}Classical phase diagram}
\begin{figure}[ht!]
\centering
\includegraphics*[clip,width=70mm]{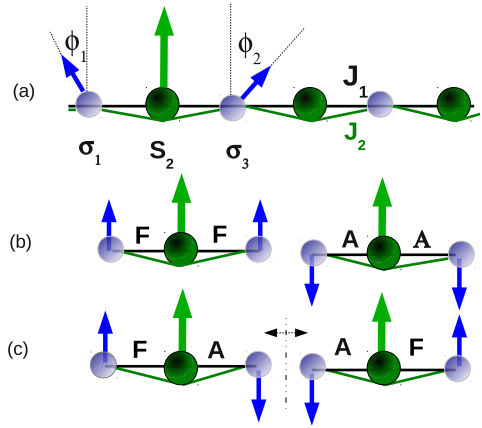}
\caption{(Color online) (a) Sketch of the
mixed-spin chain, Eq.~(\ref{h}), and the three-spin cluster used to 
construct the classical phase diagram. $\Phi_1$ and $\Phi_2$ are 
the variational parameters fixing the directions of the $\sigma$ spins.
(b,c) Four cluster ground-state  configurations used as building blocks
for  construction of the  classical phases. The doubly degenerate cluster 
state (c) suggests a $2^{L}$-fold degeneracy of the classical D phase.         
F an A stand, respectively, for FM and AFM orientations of both classical 
spins on a bond. 
} 
\label{model}
\end{figure}

%%%%%%%%%%%%%%%%%%%%%%%%%%%%%%%%%%%%%%%%%%%%%%%%%%
To establish the classical phase diagram related to  Eq.~(\ref{h}), 
it is instructive to start with an analysis of the
classical states of the local Hamiltonian $h_1$   
sketched in Fig.~\ref{model}(a). Fixing the  direction of $\bs{S}_2$, 
one finds four  different cluster spin configurations [denoted by $FF$, $AA$, $FA$, 
and $AF$ in Figs.~\ref{model}(b) and (c)] by minimizing  the cluster energy 
in the parameter regions $\frac{3\pi}{4}\leq t \leq \frac{3\pi}{2}$ ($FF$),  
$-\frac{\pi}{2}\leq t \leq \frac{\pi}{4}$ ($AA$), and  $\frac{\pi}{4}\leq t
\leq \frac{3\pi}{4}$ ($FA$ and $AF$). Here F and A mean, respectively, FM and
antiferromagnetic (AFM) orientations of the nearest-neighbor spins on a bond. 
Apart from an arbitrary global rotation of cluster spins, 
the lowest-energy state in the last sector  is  doubly degenerate,
Fig.~\ref{model}(c). 
%===================    figure   =================================
%===================    figure   =================================

The established cluster states may be used as building blocks
to construct optimal $L$-cell spin configurations, by fitting the directions of the 
sharing $\sigma$ spins of neighboring blocks. By construction, such states correspond
to local minima of the classical energy. 
Clearly, there are unique  global spin configurations constructed
only from  $FF$ or $AA$ three-spin blocks representing, respectively,  
the classical FM and (N\'{e}el-type) FiM phases. 
On the other hand, to construct the manifold  of GS's 
realized in the parameter region $\frac{\pi}{4}\leq t \leq \frac{3\pi}{4}$ 
(sector D on the phase diagram in Fig.~\ref{diagram}), 
we have to  find all  possible configurations by using  
the building blocks  $FA$ and $AF$  and their counterparts  with  opposite
spin directions. As the number of possible ways to attach  a new block to
a given global configuration is two, the degeneracy 
of the  classical ground state  in this region is exponentially large
($2^{L}$). The established classical phase diagram was additionally
confirmed by classical Monte Carlo simulations. 

Generally speaking, quantum fluctuations may be  expected to reduce the classical 
degeneracy of the D phase  and to  favor some subset of classical states. 
A peculiarity of the  three-site interaction in Eq.~(\ref{h})
is that even at a classical level it promotes only  collinear spin
configurations. As zero-point fluctuations as a rule exhibit the  same
tendency, it may be speculated  that in the quantum case the stabilized phases
will inherit this peculiarity of the classical model. In fact, the following
analysis of the quantum model confirms the above suggestion. Another special
property of the classical three-site interaction is the obvious tendency
(for $J_2>0$) towards local symmetry breaking  of the nearest-neighbor spin 
correlations. This  leads in the quantum system (see below) to a specific clustering and
the formation of  local nearest-neighbor  composite-spin
states. Finally, the systems with integer and  half-integer cell spins $S+\sigma$ 
may be  expected to exhibit different quantum phases 
in the $D$ sector. Indeed, according to the generalized Lieb-Schultz-Mattis 
theorem \cite{affleck1} (applicable to systems with  half-integer cell spins
$S+\sigma$), such  systems can have either  non-degenerate gapless GS's  or 
gapped degenerate GS's  with a broken lattice symmetry. 

%%%%%%%%%%%%%%%%%%%%%%%%%%%%%%%%%%%%%%%%%%%%%%%%%%%%%%%%%%
\section{Some exact results}
\label{sec-exact}
%%%%%%%%%%%%%%%%%%%%%%%%%%%%%%%%%%%%%%%%%%%%%%%%%%%%%%%%%%%
\subsection{One-magnon states of the FM phase}
%%%%%%%%%%%%%%%%%%%%%%%%%%%%%%%%%%%%%%%%%%%%%%%%%%%%%%%%%%%
In the alternating-spin $(S,\sigma)$ chain, there are  two types of 
one-spin-flip excitations above the fully polarized
FM state
$|F\rangle =|\sigma_1S_2\sigma_3S_4\ldots\sigma_{2L-1}S_{2L}\rangle$,
which can  be written as  
$|2m\rangle =S_{2m}^-|F\rangle$ and
$|2m-1\rangle =\sigma_{2m-1}^-|F\rangle$, where  
$S_{2m}^-=S_{2m}^x-\imath S_{2m}^y$ and   
$\sigma_{2m-1}^-=\sigma_{2m-1}^x-\imath
\sigma_{2m-1}^y$ ($m=1,2,\ldots,L$). A simple inspection of the action of 
the Hamiltonian $\cal{H}_{\sigma S}$ on these  states gives the following 
exact dispersion relations for the one-magnon excited states  above the
FM  state 
\be\label{ek}
E^{(\pm)}_k=a_k\pm\sqrt{b_k^2+c_k^2},
\ee
where $a_k=-(S+\sigma)\, \kappa-2\sigma S\sin t\, \sin^2 (ka_0)$,
 $b_k= (S-\sigma)\, \kappa+2\sigma S\sin t\, \sin^2 (ka_0)$, 
$c_k=2\sqrt{S\sigma}\, \kappa\, \cos
(ka_0)$, $\kappa = \cos t +\sigma\left(2S-1\right)\sin t$, and $a_0$ is  
the lattice spacing.   

As may be expected, there are two different types of one-magnon excitations
belonging to the gapless $E_k^{(-)}$ and optical $E_k^{(+)}$ branches.   
It is easy to check that the expressions for the instability points
$t_F$ and $t_F^{'}$  ($t_F<t^{'}_F$) of the one-magnon excitations,  
entirely determined  by  the gapless branch $E_k^{(-)}$, read
\beq\label{instability}
\cos t_{F} +2\sigma S\left(1+ \frac{1}{2S}\right)\sin t_{F}=0\nonumber \\ 
\cos t_{F}^{'} +2\sigma S\left(1- \frac{1}{2S}\right)\sin t_{F}^{'}=0.\nonumber
\eeq
In the case $(S,\sigma)=(1,\frac{1}{2})$, $t_{F}=\pi -\arctan
\left(\frac{2}{3}\right)$ and 
$t_{F}^{'}=2\pi-\arctan (2)$.
At both instability points,   $E_k^{(-)}$ softens in the whole  Brillouin zone, 
whereas  $E_k^{(+)}$ keeps its gap structure at $t=t_{F}$, but reduces to the 
gapless form $E_k^+=-4\sigma S\sin t_{F}^{'}\sin^2 (ka_0)$ at $t_{F}^{'}$.  
As proved  below, the  instability point $t_{F}$ coincides with one of 
the exact  quantum boundaries  of the FM phase in Fig.~\ref{diagram}, 
whereas $t_{F}^{'}$  is not related to  the phase boundaries.

%%%%%%%%%%%%%%%%%%%%%%%%%%%%%%%%%%%%%%%%%%%%%%%%%%%%%%%%%%%
\subsection{Three-spin cluster model}
%%%%%%%%%%%%%%%%%%%%%%%%%%%%%%%%%%%%%%%%%%%%%%%%%%%%%%%%%%%M
Some valuable information concerning  the quantum phase diagram of
Eq.~(\ref{h}) can be extracted already from the three-site cluster model  
defined by one of the local Hamiltonians in Eq.~(\ref{h}), say, $h_1$.
For $\sigma =\frac{1}{2}$,  it is instructive to recast   $h_1$ in the form
\be
h_1=J_1^{'}\bs{S}_2\cdot\bs{\sigma}_{13}+J_2\, 
\left( \bs{S}_2\cdot\bs{\sigma}_{13}\right)^2-\frac{J_2}{2}S(S+1),
\label{h1}
\ee  
where $J_1^{'}=J_1+J_2/2$ and $\bs{\sigma}_{13}=\bs{\sigma}_{1}+\bs{\sigma}_{3}$.
The spin operators $\bs{\sigma}_{13}$ and $\bs{S}_T=\bs{S}_{2}+\bs{\sigma}_{13}$
define the good quantum numbers $\sigma_{13}=0,1$ and $S_T=S\pm \frac{1}{2}$,  
which are used to classify the eigenvalues  and eigenstates of $h_1$ 
(for $S=1$, see  Table~\ref{t}). 
%=====================================================
\begin{table}[h]
\begin{center}
\begin{tabular}{l|l|l|l}
\hline
$S_T$ &$\sigma_{13}$&$\hspace{1cm}\varepsilon$&\hspace{0.3cm}Eigenstates\\
\hline
0  & 1  & $-2\cos t+2\sin t$ &
$|S\rangle =\frac{1}{\sqrt{3}} (
\xi^{+}\eta^{-}\!+\!\xi^{-}\eta^{+}$ \\
  &  & &
\phantom{$|S\rangle =$}
$-
\xi^{0}\eta^{0} )$\\[1mm]
%\hline
1& 1 & $-\cos t-\frac{1}{2}\sin t$&
$|T_1^0\rangle =\frac{1}{\sqrt{2}} (
\xi^{+}\eta^{-}\!-\!\xi^{-}\eta^{+} )$\\[1mm]
1 & 0 & $-\sin t$& $|T_2^0\rangle =\xi^{0}\eta^{s}$\\ [1mm]
2 & 1 &  $\cos t+\frac{1}{2}\sin t$&
$|Q^0\rangle =\frac{1}{\sqrt{6}} ( \xi^{+}\eta^{-}\!+\!\xi^{-}\eta^{+}$\\
 &  &  &
\phantom{$|Q^0\rangle =$}
$+ 2 \xi^{0}\eta^{0} )$\\
%\hline
\end{tabular}
\caption{
\label{t} 
The eigenvalues $\varepsilon$  and eigenstates
of the cluster Hamiltonian $h_1$ for $S=1$ and $\sigma=\frac{1}{2}$ 
in terms of the good quantum numbers  $\sigma_{13}$  and $S_T$. 
$\xi^{\mu}$ and $\eta^{\mu}$ ($\mu=0,\pm 1$) are, respectively,  the
canonical basis states of $\bs{S}_2$ and
the composite operator $\bs{\sigma}_{13}=\bs{\sigma}_{1}+\bs{\sigma}_{3}$ 
in the triplet $\sigma_{13} =1$  state.
$\eta^{s}=\frac{1}{\sqrt{2}}(\uparrow_1\downarrow_3-\downarrow_1\uparrow_3)$
is the singlet ($\sigma_{13} = 0$)  eigenstate of $\bs{\sigma}_{13}$.
For brevity, only the  $0$ components of the  triplet 
and  quintet states are presented.
}
%\vspace{-.7cm}
\end{center}
%\label{t}
\end{table}
%===========================================================
%===================    figure   =================================
\begin{figure}[ht!]
\centering
\includegraphics*[clip,width=75mm]{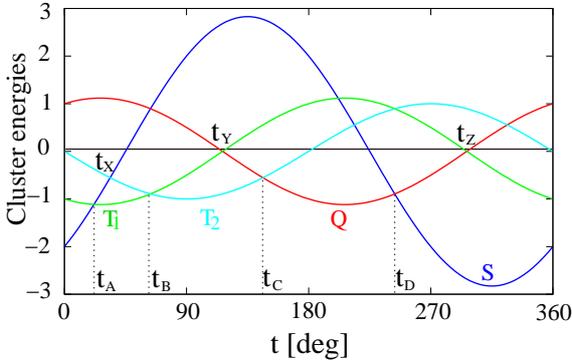}
\caption{(Color online)
Energy levels of the 3-site cluster vs. $t$. 
$S$, $T_i$ ($i=1,2$), and $Q$ are shortcuts of the singlet, triplet, and
quintet states in Table~\ref{t}.  
$t_A=\arctan\left(\frac{2}{5}\right)\approx 21.80^{\circ}$,
$t_X=\arctan\left(\frac{2}{3}\right)\approx 33.69^{\circ}$,
$t_B=\arctan\left(2\right)\approx 63.43^{\circ}$,
$t_Y=\pi-\arctan\left(2\right)\approx 116.57^{\circ}$,
$t_C=\pi-\arctan\left(\frac{2}{3}\right)\approx 146.31^{\circ}$, 
$t_D=\pi+\arctan(2)\approx 243.43^{\circ}$,  and
$t_Z=2\pi-\arctan\left(2\right)\approx 296.57^{\circ}$.
Some special properties of the  crossing points  are 
discussed in the text.
} 
\label{e_cl}
\end{figure}
%===================    figure   =================================

\begin{figure}[ht!]
\centering
\includegraphics*[clip,width=75mm]{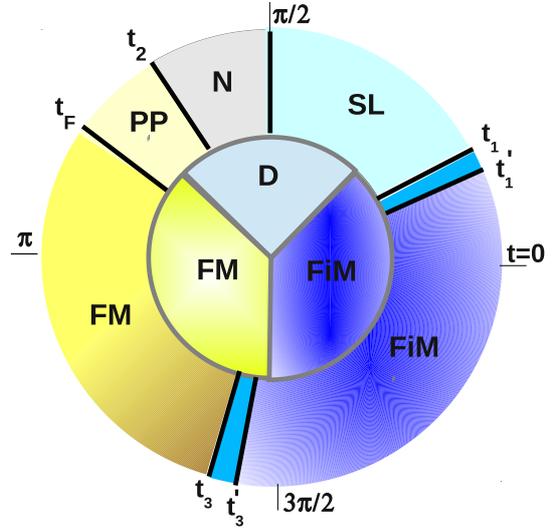}
\caption{(Color online) Classical (inner circle) and quantum (outer
circle) phase diagrams of the (1,$\frac{1}{2}$) model, Eq.~(\ref{h})
\textit{vs} $t$ ($0 \leq t < 360^{\circ}$). FM and FiM denote classical
ferromagnetic and N\'{e}el-type ferrimagnetic phases, respectively, whereas D
stands for the  classical $2^{L}$-fold degenerate phase. SL, N, and PP stand, respectively,
for  the the spin-liquid, nematic-like and partially polarized phases.
The sectors $t_1^{'}<t<t_1$   and $t_3<t<t_3^{'}$ are occupied by
intermediate partially-polarized phases described in text. $t_F=\pi-\arctan
\left(\frac{2}{3}\right)$
($\approx 146.31^{\circ}$) is an exact FM phase boundary. The DMRG estimates
for the other phase boundaries read as follows. $t_1^{'}=25.03^{\circ}$,
$t_1\simeq 30^{\circ}$, $t_2\simeq 120^{\circ}$, $t_3=253.08^{\circ}$, and
$t_3^{'}=264.0^{\circ}$.   
} 
\label{diagram}
\end{figure}

In what follows  we prove that $t_F$ is an exact phase boundary of the FM phase. To
this end,  let us firstly discuss  the structure  of the energy levels of $h_1$  
presented in Table~\ref{t} and Fig.~\ref{e_cl} for different values of the parameter $t$.  
There are four regions in the whole parameter space (separated by the crossing points 
$t_A$, $t_B$, $t_C$, and $t_D$), where the  cluster system exhibits different GS's.  
Denoting by $\varepsilon_g=\varepsilon_g(t)$ the GS energy of $h_1$, 
the cluster theorem implies that $\varepsilon_g$ serves as an  exact lower bound for the GS energy 
per cell $E_0/L$ of the quantum Hamiltonian $\cal{H}_{\sigma S}$. 
Since  the energy of the quintet state $|Q^0\rangle$ coincides with the energy per site of the 
FM phase (see Table~\ref{t}), we conclude that the FM state  is the GS  of Eq.~(\ref{h})
 in the region  $t_C\leq t\leq t_D$,  where $|Q^0\rangle$ is a
cluster GS. Moreover, since  the FM phase  is gapless, the generalized 
Lieb-Schultz-Mattis theorem implies that there are no other GS's. 
Finally, since $t_C$ coincides with  the  one-magnon instability point
$t_F$ of the FM phase, we conclude that $t_F$ is also an exact quantum phase boundary 
of the FM phase.    Notice that the other boundary $t_D$ of the  quintet state 
can not be directly related to the other   $FM$ phase boundary $t_3$
in Fig.~\ref{diagram}  because the  one-magnon instability point $t_F^{'}$  
lies beyond the region $t_C<t<t_D$.   As a matter of fact, the  DMRG
estimate  $t_3\approx 253.08^{\circ}$  implies that $t_D<t_3<t_Z$, whereas
the instability point $t_F^{'}$  coincides with the  crossing point $t_Z$.

 The established  connection  between crossing points  in 
Fig.~\ref{e_cl} and some special points on the quantum phase diagram, Fig.~\ref{diagram},
  can be further extended as follows.\\
(i) $t_X$:  At this point, $h_1$ is recast to the 
form $h_1=J_2\bar{h_1}-J_2-J_2 S(S+1)/2$, where  
$\bar{h_1}=\left(1+\bs{S}_{2}\cdot\bs{\sigma}_{13}\right)^2$ and $J_2>0$.
Since $\bar{h_1}|T_1^{\mu}\rangle=0$, the numerical DMRG estimate 
for the GS matrix element  $\langle \bar{h_1}\rangle\approx
0.006$  implies that the GS  at $t=t_X$ is predominantly 
constructed from local spin configurations related to the triplet cluster
states  $|T_1^{\mu}\rangle$.\\
(ii) $t_B$ ($J_1^{'}=J_2>0$): This point appears in the middle 
of the non-magnetic region in Fig.~\ref{diagram}. At $t=t_B$, the cluster 
Hamiltonian  $h_1$ is   proportional to the projector  operator 
${\cal P}_1=1-\frac{1}{2}\bs{S}_{2}\cdot\bs{\sigma}_{13}-\frac{1}{2} 
\left( \bs{S}_{2}\cdot\bs{\sigma}_{13}\right)^2$ projecting 
onto  the subspace  spanned  by  the triplet states
$|T_1^{\mu}\rangle$ and  $|T_2^{\mu}\rangle$. 
In terms of the projectors  ${\cal Q}_n=1-{\cal P}_n$,    
$$
{\cal H}_{\sigma S}=2J_2\sum_{n=1}^L{\cal Q}_n-J_2L,\hspace{0.2cm} J_2>0. 
$$
This means  that at $t=t_B$ the GS of ${\cal H}_{\sigma S}$ may be sought 
as an optimal product state composed of local triplet states 
($|T_1^{\mu}\rangle$ and  $|T_2^{\mu}\rangle$).\\  
(iii) $t_Y$ ($J_1^{'}=0$, $J_2>0$):  Relatively close to this point 
(around $t=t_2$) there are pronounced changes of the short-ranged
(SR) correlator  $\langle\bs{S}_{2n}\cdot\bs{S}_{2n+2}\rangle$, indicating 
a quantum  phase transition between the magnetic PP and non-magnetic N 
phases in Fig.~\ref{diagram}.
%===================    figure   =================================
%===================    figure   =================================

%%%%%%%%%%%%%%%%%%%%%%%%%%%%%%%%%%%%%%%%%%%%%%%%%%%%%%%%%%%%%%%%%
\section{\label{sec_qpd} Quantum phase diagram}
%%%%%%%%%%%%%%%%%%%%%%%%%%%%%%%%%%%%%%%%%%%%%%%%%%%%%%%%%%%%%%%%%%
Most of the the numerical results in this Section are obtained by DMRG simulations
and concern properties of the quantum phase diagram of model~(\ref{h}) in 
the extreme quantum case $(S,\sigma)=(1,\frac{1}{2})$, Fig.~\ref{diagram} (outer circle). 
As a rule, there have been  performed 7 DMRG sweeps, keeping up
to 500 states in the last sweep.  
The above conditions ensure a good convergence up to 256 unit cells, with a discarded
weight of the order of $10^{-8}$ or better. 
%===================    figure   =================================
\begin{figure}
\includegraphics*[clip,width=75mm]{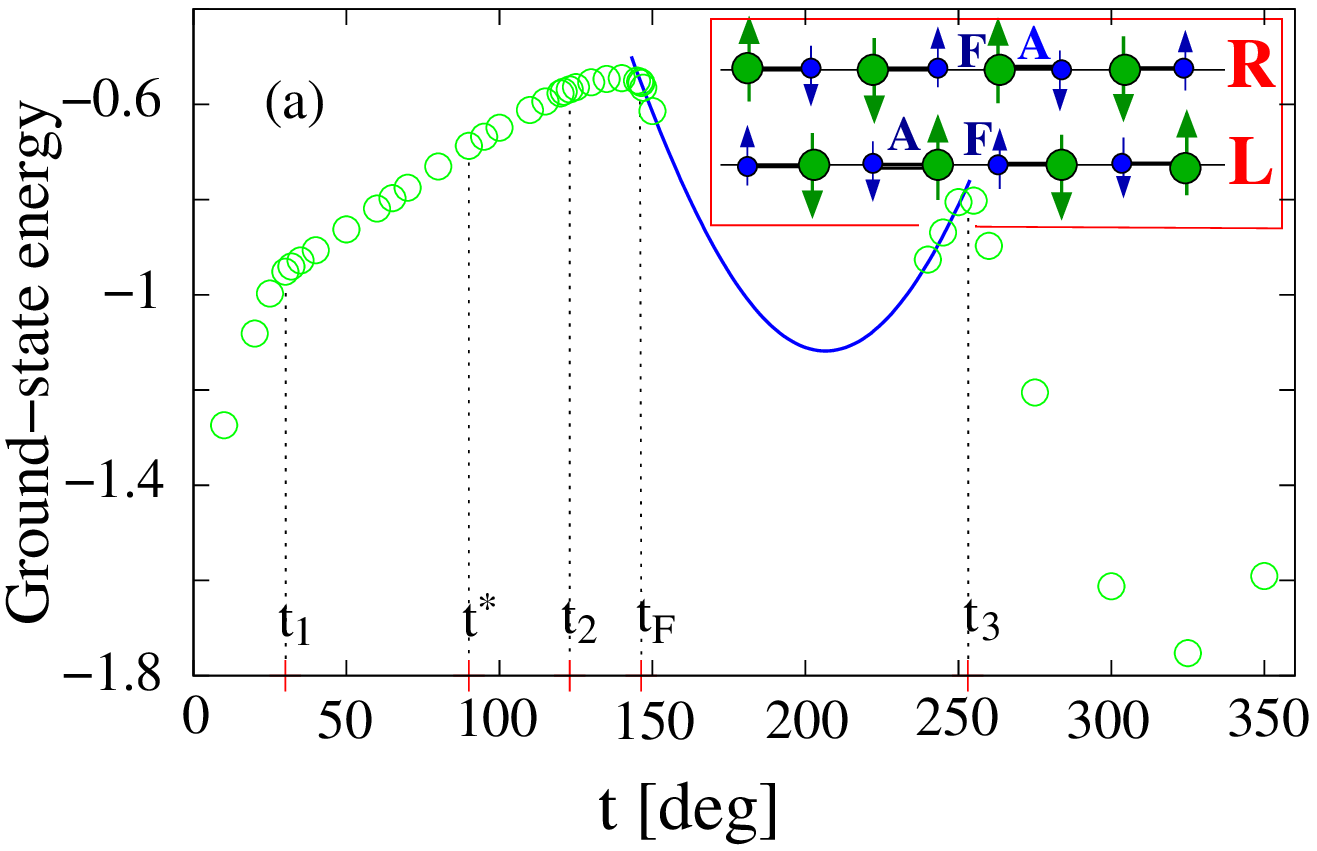}
\includegraphics*[clip,width=75mm]{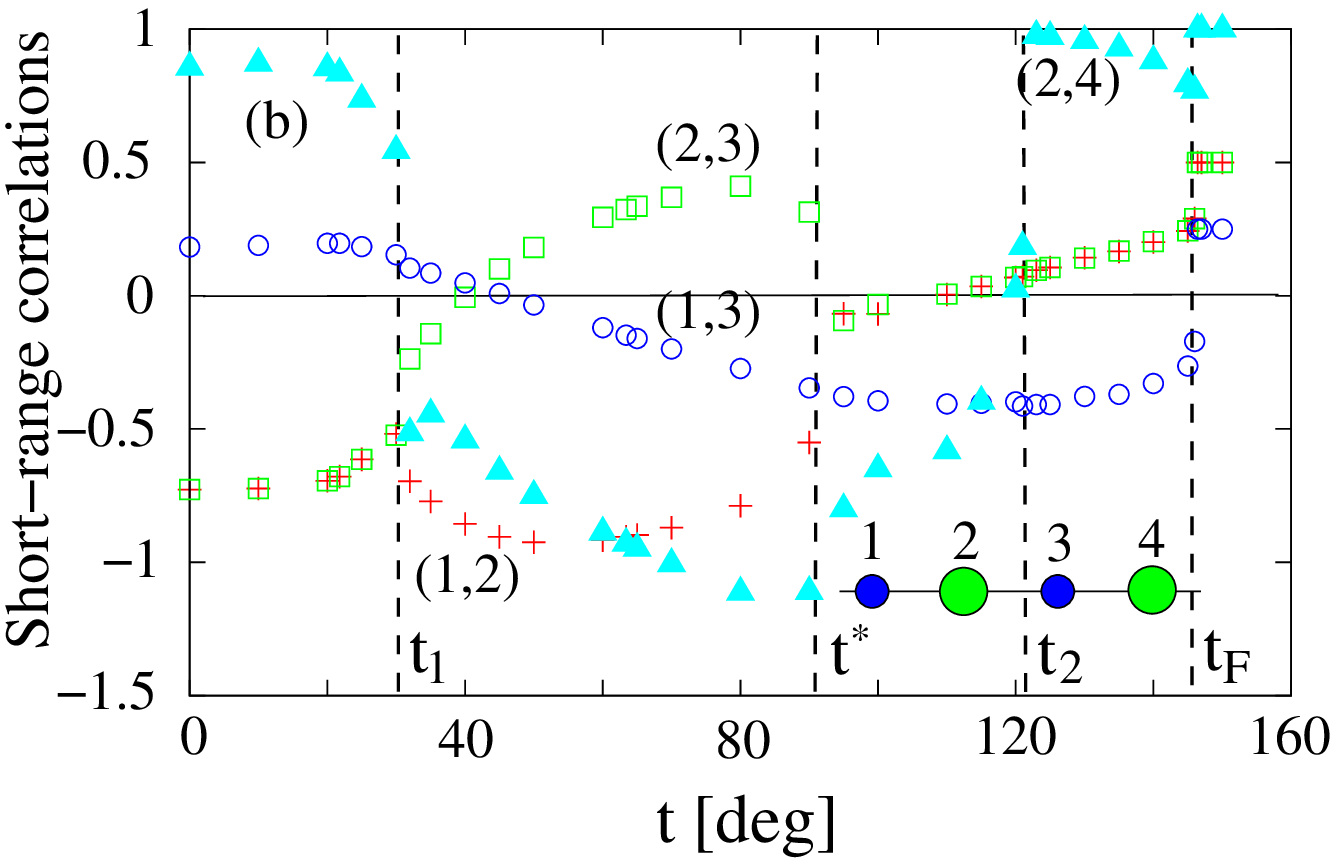}
\caption{(Color online) (a) DMRG results (OBC, $L=100$) for the GS energy per cell
of the   $(1,\frac{1}{2})$ chain as function of  $t$. The line shows the exact 
GS energy of the FM state. $t_1$, $t^*\simeq \frac{\pi}{2}$, $t_2$, $t_F$,  and $t_3$ are the
phase transition points displayed in  Fig.~\ref{diagram}.
Inset: The cluster configurations $|\Psi_L\rangle$ and
$|\Psi_R\rangle$ selected by the two types of  OBC. 
(b) DMRG results for some SR spin-spin correlators of the   $(1,\frac{1}{2})$ 
chain as a function of  $t$.  The symbols  $(i,j)$ are shortcuts of the isotropic
spin-spin correlators  between the spins at 
sites $i$ and $j$ (OBC, $L=100$). 
} 
\label{src}
\end{figure}
%%===================    figure   =================================

To begin with, let us discuss the general structure and some peculiarities 
of the quantum phase diagram, Fig.~\ref{diagram}, related to the three-body 
exchange interaction. Some  features  of the diagram  are  encoded  in the behavior of the 
SR correlators presented in Fig.~\ref{src}. As may be expected, the most complex  
behavior (with abrupt changes of the SR correlators) appears in the region 
characterized by a manifold of degenerate classical GS configurations 
(D sector in  Fig.~\ref{diagram}). As argued below, the abrupt 
changes of the  SR correlator $\langle \bs{S}_{2n}\cdot\bs{S}_{2n+2}\rangle$ 
around the points  $t_F$ and $t_2$ are related with  the emergence 
of partially-polarized states 
mediating the transition from  the FM to a non-magnetic (N) state. 
In fact, it occurred that the destruction of both classical magnetic
phases (FM and FiM) takes place only through intermediate (partially-polarized) 
states, located in the sectors $(t_1^{'},t_1)$,  $(t_2,t_F)$ and $(t_3,t_3^{'})$
in Fig.~\ref{diagram}.

The mentioned  clustering  effect of the three-body interaction is 
characteristic for the (non-magnetic) SL sector  in Fig.~\ref{diagram} and 
suggests an establishment of the alternating-bond GS structure 
$uvuv\ldots$ with  $u\neq v$, where
$u=\langle\bs{S}_{2n}\cdot\bs{\sigma}_{2n-1}\rangle$ and
$v=\langle\bs{S}_{2n}\cdot\bs{\sigma}_{2n+1}\rangle$ ($n=1,\ldots,L$). 
Clearly, in the periodic chain there are two
equivalent types of "dimerized" states 
[denoted by  $|\Psi_L\rangle$ and
$|\Psi_R\rangle$ in the Inset of Fig.~\ref{src}(a)], 
which correspond to
both types of clustering  ($uv$ and $vu$) of the local Hamiltonians $h_n$
in Eq.~(\ref{h}). The clustering effect is strongly pronounced 
especially in the middle of the SL region (i.e., close to  the  
point $t_B\approx 63.43^o$, Fig.~\ref{e_cl}), where the values $(u,v)\approx
(-1,\frac{1}{3})$, red crosses and green squares in Fig.~\ref{src}~(b), indicate 
the formation of almost pure  spin-$\frac{1}{2}$ states of the composite
cell spin $\bs{S}_{2s}+\bs{\sigma}_{2n-1}$.   $|\Psi_L\rangle$ and
$|\Psi_R\rangle$ are related by the site parity operation
${\cal P}|\Psi_{L,R}\rangle=|\Psi_{R,L}\rangle$.  It is important to emphasize  that
the established clustering does not violate the  original translation
symmetry by two lattice sites of the Hamiltonian (\ref{h}). 
As demonstrated in the Inset of Fig.~\ref{src}(a), the  two types of cluster states can 
be stabilized by using two different types of open boundary conditions (OBC).
%===================    figure   =================================
\begin{figure}
\includegraphics*[clip,width=75mm]{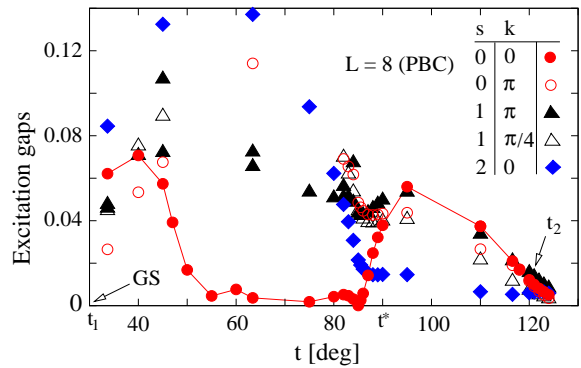}
\caption{
(Color online) Numerical exact diagonalization (ED) results for the low-lying excitation
gaps in  the periodic $L=8$ $(1,\frac{1}{2})$ system, Eq.~(\ref{h}),  as functions of the
parameter $t$. Inset: spins ($s$) and wave vectors ($k$) of
the excitations. In the region $t_1\lesssim t\lesssim t^*$, the lowest
excited state is a singlet, which scales exponentially with $L$ to the GS. 
Close to  $t=t^*\approx \frac{\pi}{2}$  the quintet 
excitation ($s=2$) is soften and becomes the lowest 
excitation for $t>t^*$.  
} 
\label{gaps}
\end{figure}
%===================    figure   =================================
In finite periodic  chains, the symmetry $\cal P$ is not violated,
so that we can  expect two lowest quasi-degenerate states related to the 
symmetric (antisymmetric) combinations 
$|\Psi_{\pm}\rangle=\frac{1}{\sqrt{2}}\left(|\Psi_{L}\rangle\pm |\Psi_{R}\rangle\right)$.  
As seen  in Fig.~\ref{gaps}, the expected structure  of the spectrum with a 
singlet lowest-lying excitation is revealed even  in small rings.         
In fact, the performed finite-scaling scaling (FSS) analysis -- using additional 
DMRG results for larger periodic
systems -- implies that in the SL region the lowest 
singlet excitation  scales exponentially fast with $L$ to 
the singlet GS, the characteristic length being strongly dependent on the parameter
 $t$. As discussed below, such a doubling of the spectrum can be identified as
well for triplet lowest-lying states.  Finally, the numerical ED results
 presented in Fig.~\ref{gaps} imply a different structure
 of the states in the non-magnetic region $t^*<t<t_2$  
which is characterized by a  quintet ($s=2$) lowest-lying excited state. 
%%%%%%%%%%%%%%%%%%%%%%%%%%%%%%%%%%%%%%%%%%%%%%%%%%%%%%%%%%%%%%%%%%%%%%%%%%%%%%%%%
\subsection{\label{sec:} Intermediate  magnetic states}
%%%%%%%%%%%%%%%%%%%%%%%%%%%%%%%%%%%%%%%%%%%%%%%%%%%%%%%%%%%%%%%%%%%%%%%%%%%%%%%%%
In this Subsection, we discuss properties   of the  intermediate partially-polarized 
magnetic states identified close to the boundaries of the classical FM and FiM
phases in the sectors PP, $t_1^{'}<t<t_1$, and $t_3<t<t_3^{'}$ 
(see Fig.~\ref{diagram}). These states do not appear in the 
classical phase diagram.\\

\begin{figure}
\centering
\includegraphics*[clip,angle=270,width=70mm]{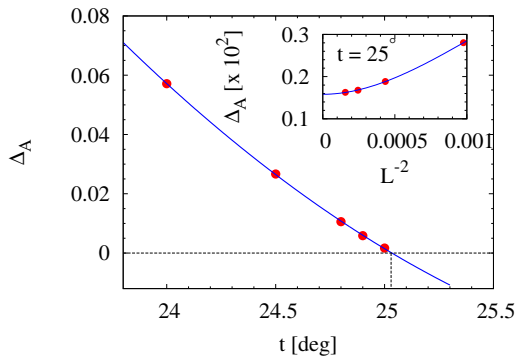}
\caption{(Color online)
DMRG results (red filled circles: $L=64$, PBC) for the AFM gap 
$\Delta_A$ \textit{vs} $t$ of the FiM phase close to the point 
$t_1^{'}$ where $\Delta_A (t_1^{'})=0$. The solid line represents the 
fit to the data obtained by the three-parameter ansatz
$\Delta_A(t)=b_0+b_1t+b_2t^2$. The Inset shows the FSS
 data for  $\Delta_A(L)$ at $t=25^{\circ}$ (red filled circles) 
and the fit (solid line) obtained by the ansatz
$\Delta_A (L)=\Delta_A (\infty)+a_1/L^4+a_2/L^6$ (solid line),
where $\Delta_A(\infty)=0.00158$. 
} 
\label{DA24_25}
\end{figure}

%%%%%%%%%%%%%%%%%%%%%%%%%%%%%%%%%%%%%%%%%%%%%%%%%%%%%%%%%%%%%%%%%%%%%%%%%%
(i) \textit{Magnetic states in the the sector $t_1^{'}<t<t_1$}:

Denoting by $E(M)$ the lowest-energy eigenvalue in the subspace defined by
the  z component of the total spin $M$, the gap  of  the
one-magnon AFM branch of excitations in the FiM phase reads
$\Delta_A=E(M+1)-E(M)$. Here $M=(S-\sigma)L=L/2$ defines the  
GS sector  of the  Lieb-Mattis-type  FiM phase characterized
by the cell magnetic  moment  $m_0\equiv M/L=\frac{1}{2}$. 
A  major  effect of the competing three-body interaction in Eq.~(\ref{h}) 
is the monotonic reduction of the gap with $t$ in the whole 
$t>0$ region,  up to the point  $t_1^{'}=25.03^{\circ}$ 
where $\Delta_A$  vanishes, 
Fig.~\ref{DA24_25}. In the same interval, the local magnetic moments 
$m_S=\langle S_{2n}^z\rangle$  and $m_{\sigma}=\langle \sigma_{2n-1}^z\rangle$ 
exhibit non-monotonic behavior. In particular, they reach their extremal values 
(maximum and minimum, respectively)  around one of the crossing points of the cluster
model, namely,  $t_A=\arctan (\frac{2}{5})\approx 21.8^{\circ}$ \cite{AKLT}. 
The magnetic moments remain finite at the critical point  $t_1^{'}$ ($m_S=0.78$ and
$m_{\sigma}=0.28$).  As may be expected, the quantized magnetization
$m_0=\frac{1}{2}$ --
%???
 а characteristic property of  the Lieb-Mattis type 
phases -- remains unchanged in the whole FiM region in Fig.~\ref{diagram}, 
up to the transition point $t_1^{'}$.
%--------------------------------------------------
%-----------------------------------------

%------------------------------------------
\begin{figure}
\begin{minipage}[b]{0.45\linewidth}
\centering
\includegraphics*[clip,width=40mm]{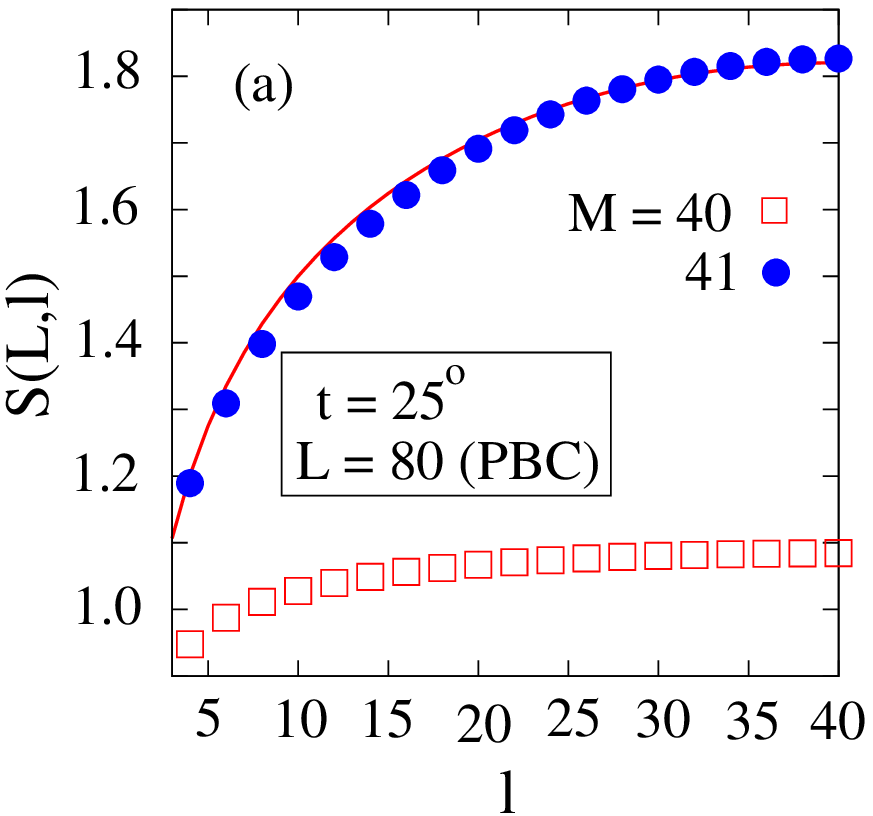}
\end{minipage}
\hspace{0.2cm}
\begin{minipage}[b]{0.45\linewidth}
\centering
\includegraphics*[clip,width=40mm]{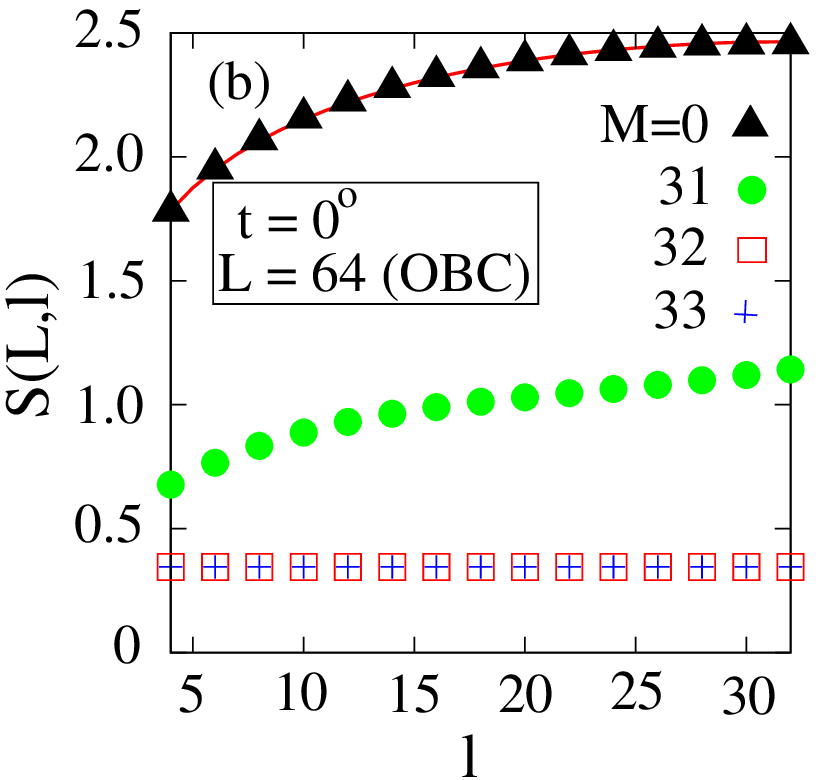}
\end{minipage}
\caption{(Color online)
DMRG results for the entanglement entropy of different states of 
the  $(1,\frac{1}{2})$  system [(a) $t=25^{\circ}$  and (b) $t=0^{\circ}$] 
as function  of the number of subblock cells $l$.  
The solid lines represent  the analytical result, Eq.~(\ref{EE}), 
for $(\eta, c)=(1,1)$ and  $(\eta, c) =(2,2.5)$
in  the first and second plots, respectively. The lowest-energy ($M=\frac{L}{2}$) state  
corresponds to the GS of a  Lieb-Mattis-type  FiM phase. The
lowest-energy ($M=\frac{L}{2}+1$) excited states,  corresponding  to one-magnon AFM
excitations, show different entropy behaviors, approximately corresponding
to central charges $c=0$ and $1$ of the system at $t=0^{\circ}$ and $25^{\circ}$, respectively.
} 
\label{ent_t0-25}
\end{figure}
%----------------------------------------

The  effect of the three-body interaction 
is reminiscent of the effect of an  applied magnetic 
field in Heisenberg ferrimagnets. A strong 
magnetic field closes  the gap  $\Delta_A$ and drives 
the system into  a  Luttinger-liquid-type magnetic 
state, which is characterized by a critical
AFM mode and a gapped  low-lying FM branch of 
excitations.  However, since the three-body interaction 
does not violate spin rotation symmetries  of the Hamiltonian, 
both interactions  might produce different states.
An interesting example is the spontaneously magnetized 
Luttinger-liquid state with gapless AFM and 
FM branches of excitations predicted for a number
of frustrated 1D ferrimagnetic systems \cite{smtll}.

In Fig.~\ref{ent_t0-25}, we present  DMRG results for 
the entanglement entropies  $S(L,l)$ of different 
low-lying states   at $t=0^{\circ}$ and $25^{\circ}$.
The well-known  analytical result for the GS  
entanglement entropy in critical
conformally-invariant 1D systems
reads \cite{holzhey}
\be\label{EE}
S(L,l)=\frac{c}{3\eta}\ln\left[ \frac{\eta L}{\pi}\sin\left(\frac{\pi
l}{L}\right)\right]+\mathrm{const}.
\ee   
Here $l$ is the number of unit cells in the subblock ($l=1,\ldots,L$),
$c$ is the central charge, and $\eta=1,2$ for PBC and OBC, respectively. 
A remarkable  fact is that the above analytical 
expression is also applicable in the case of some pure excited
states that correspond to primary fields in conformal field
theory \cite{alcaraz}.
Figure~\ref{ent_t0-25}(a) demonstrates  that  the entanglement 
entropy of the lowest-energy state in the $M+1$ sector 
closely follows the analytical expression for $c=1$. 
Thus, it can be suggested that at 
the critical point $t_1^{'}$ the system is 
spontaneously driven into a gapless 
Luttinger-liquid-type magnetic state. 
For comparison, the same excited state in the 
unfrustrated model ($J_2=0$) exhibits a constant 
entropy corresponding to the central charge $c=0$, 
Fig.~\ref{ent_t0-25}(b). 
Note also the curious observation that  the entanglement
 entropy of the lowest-energy  
$M=0$ state  of the unfrustrated ferrimagnet perfectly 
reproduces    the analytical result in  
Eq.~(\ref{EE}) for  $c=\frac{5}{2}$.    
%---------------------------------------------------
\begin{figure}
\begin{minipage}[b]{0.45\linewidth}
\centering
\includegraphics*[clip,width=41mm]{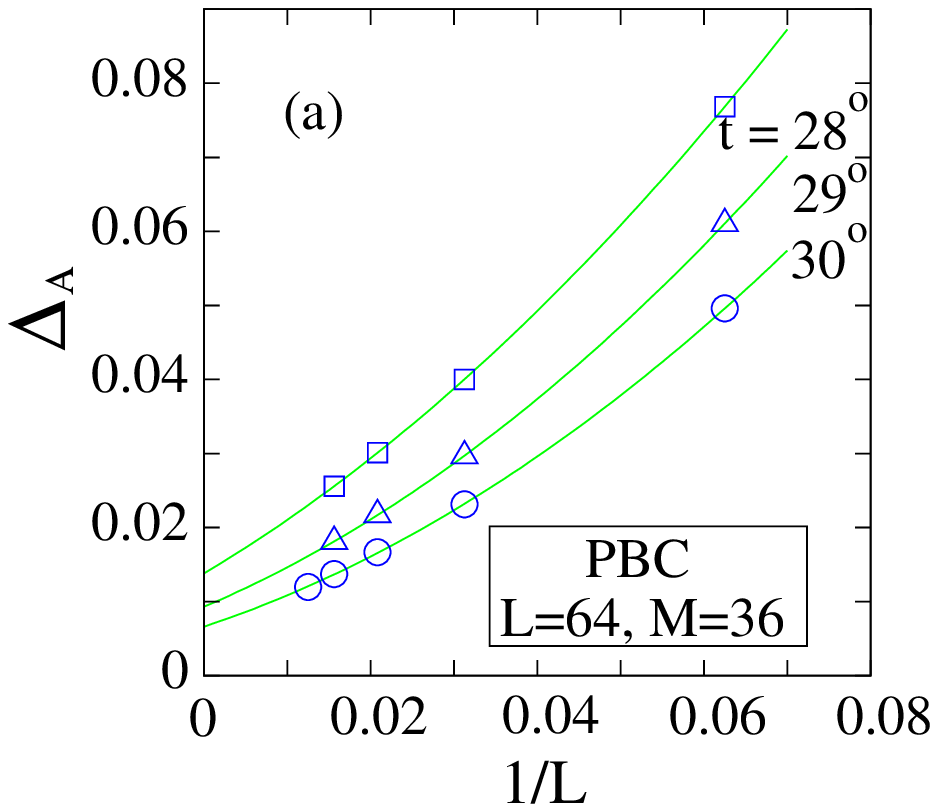}
\end{minipage}
\hspace{0.2cm}
\begin{minipage}[b]{0.45\linewidth}
\centering
\includegraphics*[clip,width=40mm]{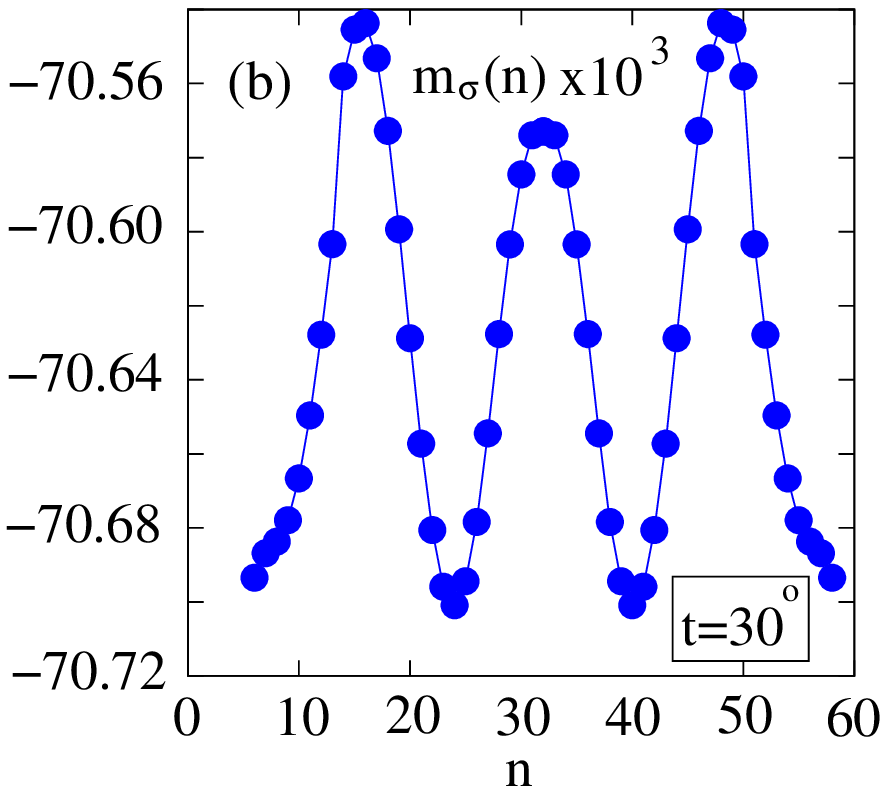}
\end{minipage}
\caption{
(Color online) (a) DMRG results for the finite-size scaling of the 
AFM gaps $\Delta_A=E(M+1)-E(M)$  in the plateaux state with magnetization $m_0=\frac{9}{16}$.
The solid lines represent the optimized  least-squares fitting
functions $\Delta_A=c_0+c_1/L+c_2/L^2$. The
extrapolated gap for $t=30^{\circ}$ is $\Delta_A=0.0066$; (b) The
local magnetic moment $m_{\sigma}(n)=\langle \sigma_{2n-1}^z\rangle$
\textit{vs} the cell index $n$ in the plateaux phase $m_0=\frac{9}{16}$
($t=30^{\circ}$, $L=64$, PBC). The extremal values of $m_{\sigma}(n)$
correspond to  the period $q=16$.
} 
\label{DA28_30}
\end{figure}
%------------------------------------------------------

A detailed  description of the magnetic phase(s)  
in the whole interval
 $t_1^{'}<t<t_1$  requires more extensive 
calculations. For instance, to obtain  the GS magnetic 
moment $M$ -- defined as the largest $M$ number with the 
property $E(M)=E(0)$ -- we need a series of  
lowest-energy eigenvalues $E(M)$ with increasing $M$. 
In fact, such  DMRG calculations  were performed for 
a few points in the above interval, including the 
boundary point $t=30^{\circ}$ which is supposed to lie 
close to the phase transition point.   The numerical  
results (up to $L=64$, PBC) imply a very slight monotonic 
increase of the GS magnetization $m_0$ from $\frac{1}{2}$ 
($t=t_1^{'}$) approximately up to the value
$\frac{9}{16}$ at $t=30^{\circ}$. The increase of $m_0$
results from a reduction of the (averaged over cells)  
magnetic moment  $\langle m_S(n)\rangle$ from $0.79$ 
($t=25^{\circ}$)  down to $0.63$ ($t=30^{\circ}$). 
In the same interval, the magnetic moment  
$\langle m_{\sigma}(n)\rangle$ increases 
from $-0.29$ up  to the value  $-0.07$.
 The abrupt change of the correlations
$\langle \bs{S}_{2n}\cdot\bs{S}_{2n+2}\rangle$ in 
the vicinity of $t=30^{\circ}$, Fig.~\ref{src}(b), 
suggests a sharp transition to the  non-magnetic state.  
According to the   general rule \cite{oshikawa}
\be\label{rule}
q\left(S+\sigma-m_0\right)=\mathrm{integer},
\ee 
the magnetization $m_0=\frac{9}{16}$ may be related
to a gapped plateau  phase characterized by 
a periodic magnetic structure with a  period $q=16$ 
unit cells.  
As a matter of fact, the numerical results for 
$m_{\sigma}(n)$, 
Fig.~\ref{DA28_30}(b), reveal such a periodic  structure,
albeit  with extremely  small  amplitudes of magnetic 
oscillations. DMRG estimates for the 
AFM gap $\Delta_A$ of  the $m_0=\frac{9}{16}$ state, 
Fig.~\ref{DA28_30}(a), imply a smooth reduction of
$\Delta_A$ with $t$ from $0.0160$ ($t=28^{\circ}$) 
down to the value $0.0066$ ($t=30^{\circ}$). 
Unfortunately, due to strong boundary 
effects -- resulting from the extreme smallness 
of the local  magnetic moments 
$m_{\sigma}(n)$ -- the suggested plateau state  can not be 
definitely established by larger-scale  DMRG calculations  
under OBC.  Thus, it may be speculated that the 
spontaneously magnetized Luttinger-liquid state 
established at $t=25.03^{\circ}$ survives up to the 
transition point about $t\approx 30^{\circ}$, 
although the  numerical results can not definitely 
exclude the scenario with some intermediate 
plateaux states in the region  $t_1^{'}<t<t_1$.
%-----------------------------------------
\begin{figure}
\begin{minipage}[b]{0.45\linewidth}
\centering
\includegraphics*[clip,width=40mm]{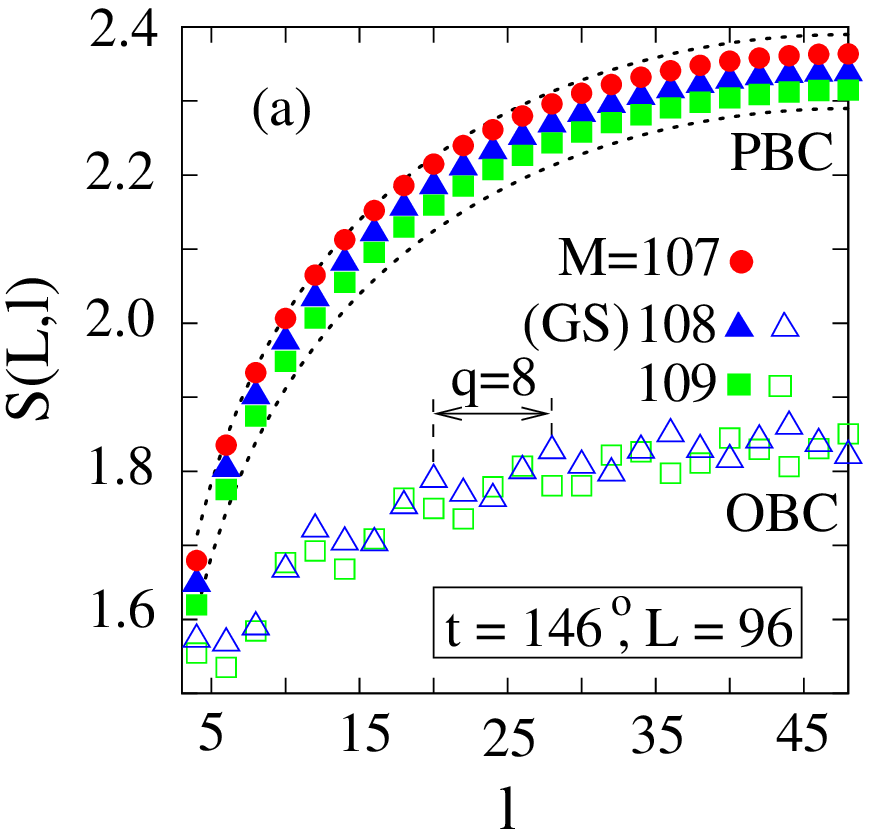}
\end{minipage}
\hspace{0.2cm}
\begin{minipage}[b]{0.45\linewidth}
\centering
\includegraphics*[clip,width=40mm]{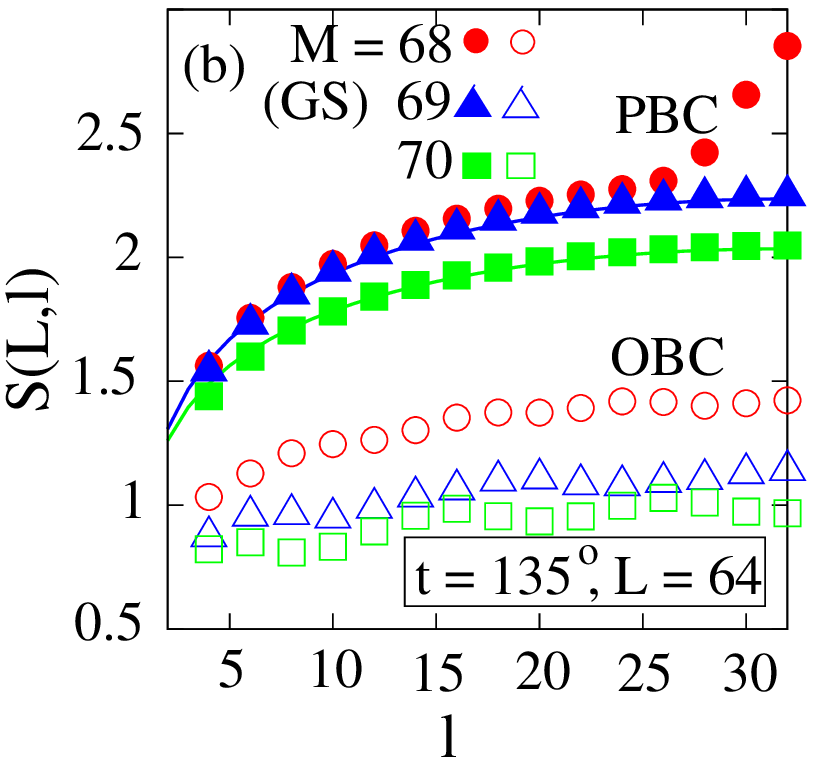}
\end{minipage}
\caption{(Color online)
DMRG results [filled (PBC) and open (OBC) symbols] for  the entanglement
 entropies $S(L,l)$ \textit{vs} $l$  of some low-lying states
of the $(1,\frac{1}{2})$ model  at (a) $t=146^{\circ}$ and (b)
$t=135^{\circ}$. Lines: the theoretical result for $S(L,l)$,
Eq.~(\ref{EE}),  with central charge $c=1$. DMRG results imply the GS magnetizations
$m_0=\frac{9}{8}=1.125$ and $\frac{69}{64}\approx 1.078$ at $t=146^{\circ}$ and $135^{\circ}$,
respectively. Under OBC, the function  $S(L,l)$ exhibits   periodic
structures ($q=8$ and $10$) at $t=146^{\circ}$ and $135^{\circ}$,
respectively.       
} 
\label{Ent135_146}
\end{figure}
%%%%%%%%%%%%%%%%%%%%%%%%%%%%%%%%%%%%%%%%%%%%%%%%%%%%%

(ii) \textit{Magnetic states in the PP sector}:\\
As argued above, the  exact phase boundary  $t_F$ coincides with one of
the instability points of the one-magnon AFM excitations and is
characterized by a complete softening of the dispersion function $E_k^{(-)}$, Eq.~(\ref{ek}),
in the whole Brillouin zone. The transition at $t_F$ is signaled by 
sharp reconstructions of  the SR correlations, 
the jump (with a change of sign) of the correlator
$\langle\bs{\sigma}_{2n-1}\cdot\bs{\sigma}_{2n+1}\rangle$
being the most important.  On the contrary, the 
nearest-neighbor correlator 
$\langle\bs{S}_{2n}\cdot\bs{S}_{2n+2}\rangle$ remains 
positive and signals a FM ordering of the spin-$S$ subsystem 
in the entire PP sector in  Fig.~\ref{diagram}. 
Close to the other boundary  $t_2$ ($\simeq \frac{2\pi}{3}$), 
the behavior of the nearest-neighbor spin correlations 
is reversed, namely, the transition to a non-magnetic state 
is accompanied by an abrupt change of sign of the correlator 
$\langle \bs{S}_{2n}\cdot\bs{S}_{2n+2}\rangle$, 
whereas the nearest-neighbor $\sigma$-spin correlations 
remain almost untouched. 
In fact, the  numerical DMRG analysis implies finite 
sublattice magnetizations [$m_S(n),m_{\sigma}(n)>0$] 
all over the  region $t_2<t<t_F$. Moreover, while the 
average of $m_{\sigma}(n)$ monotonically decreases from 
$0.221$ at $t=146^{\circ}$ down to $\approx 0.06$ at 
$t=123^{\circ}$, the average of $m_S(n)$ increases from 
$0.903$ (at $t=146^{\circ}$) almost  to its saturation 
value $1$ in a vicinity of $t_2$, where the correlations 
between $S$ and $\sigma$ sublattice spins become extremely 
weak (see Fig.~\ref{src}) and then sharply drop
 to zero.       

\begin{figure}
\begin{minipage}[b]{0.45\linewidth}
\centering
\includegraphics*[clip,width=40mm]{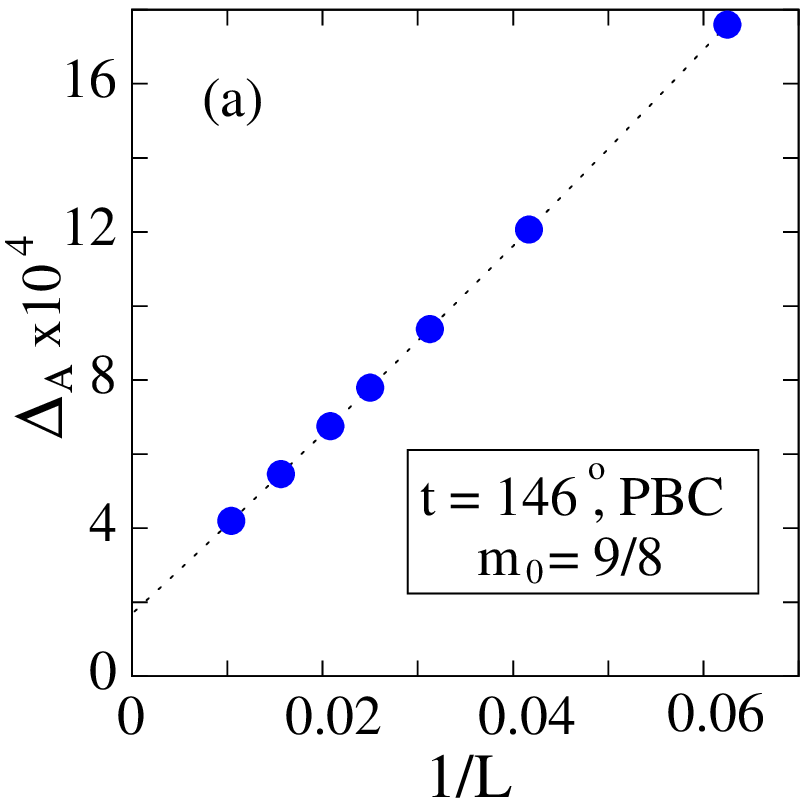}
\end{minipage}
\hspace{0.2cm}
\begin{minipage}[b]{0.45\linewidth}
\centering
\includegraphics*[clip,width=40mm]{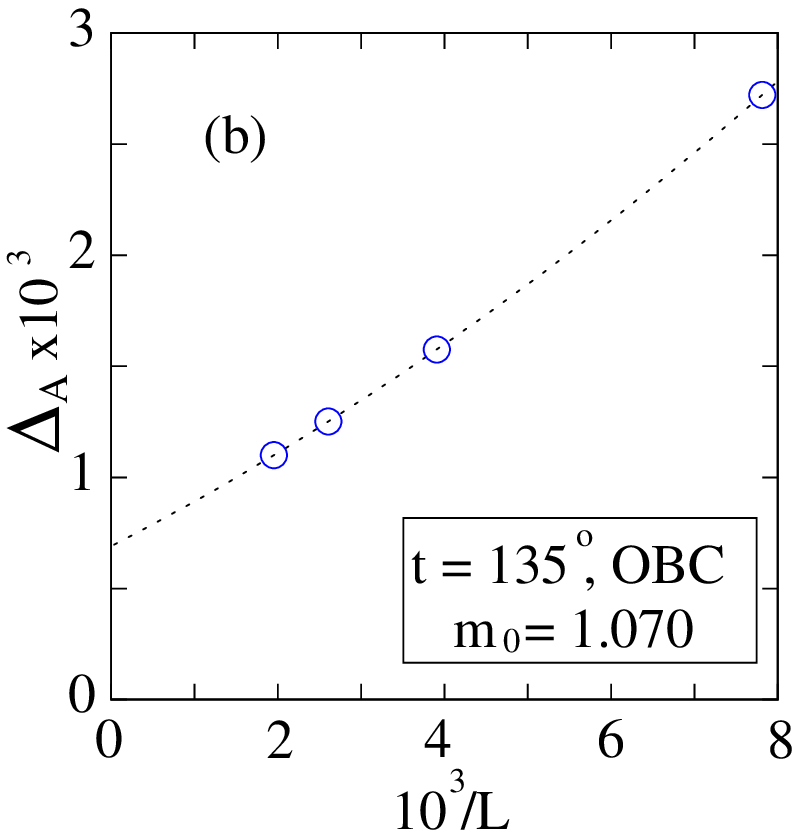}
\end{minipage}
\caption{(Color online)
Finite-size scaling of the lowest  excitation gaps in the states with GS
magnetizations (a) $m_0=\frac{9}{8}$ ($t=146^{\circ}$) and (b) $m_0=1.070$
($t=135^{\circ}$). The respective $L=\infty$ gaps $\Delta_A=1.7$ x $10^{-4}$
and $6.9$ x $10^{-4}$ are extracted from the fitting ansatz
$\Delta_A(L)=\Delta_A+a_1/L+a_2/L^2$ (dot lines).
} 
\label{DA}
\end{figure}

We show in Fig.~\ref{Ent135_146}  DMRG results for the 
entanglement entropies of a few low-lying states at two 
points ($t=146^{\circ}$ and $135^{\circ}$) corresponding 
to the GS  magnetizations $m_0=\frac{9}{8}$ and 
$m_0\approx 1.07$, respectively. 
The GS entropies at both points 
approximately follow the theoretical curves corresponding
to the central charge $c=1$. The same is true for the 
lowest-energy states in the neighbor sectors
$M\pm 1$ ($M=108$) at $t=146^{\circ}$, 
whereas at $t=135^{\circ}$ the lowest-energy states in the 
neighbor $M$ sectors apparently deviate from the
theoretical $c=1$ curve. Since an analysis based alone on 
the entanglement entropy can not definitely exclude the 
scenario with extremely small  gaps, we have
performed a separate DMRG test of the AFM gaps $\Delta_A$ 
at both points. The numerical data for $\Delta_A(L)$  
at fixed GS magnetizations $m_0$, Fig.~\ref{DA}, implies 
extremely small (but finite) extrapolated gaps at 
both points:  
$\Delta_A=1.7$ x $10^{-4}$ ($6.9$ x $10^{-4}$) at $t=146^{\circ}$
($135^{\circ}$).              
These observations resemble  the established picture  of 
magnetic states close to the transition point
$t_1$.  As before, it may be suggested that  close 
to $t_F$ a plateau $m_0=\frac{9}{8}$ state with the 
period $q=8$ is established. Unlike the 
state around $t_1$, the plateau GS  with 
$m_0=\frac{9}{8}$ around $t_F$ 
is additionally supported by   larger-scale 
DMRG results under OBC (up to $L=512$).  
Notice that the observed  critical behavior of some 
excited states  
in Fig.~\ref{Ent135_146}(a) close to the 
transition point $t_F$  is compatible with the  
established complete softening of the dispersion function 
$E_k^{(-)}$ at $t=t_F$.  
%--------------------------------------------------------------
%--------------------------------------------
\begin{figure}
\begin{minipage}[b]{0.45\linewidth}
\centering
\includegraphics*[clip,width=40mm]{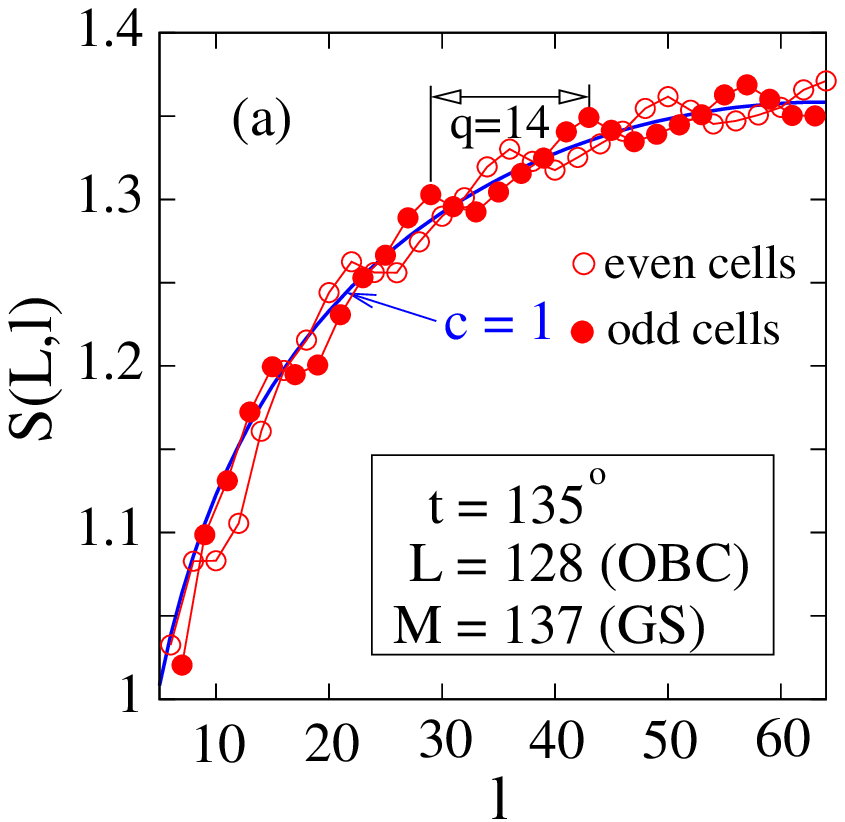}
\end{minipage}
\hspace{0.2cm}
\begin{minipage}[b]{0.45\linewidth}
\centering
\includegraphics*[clip,width=41mm]{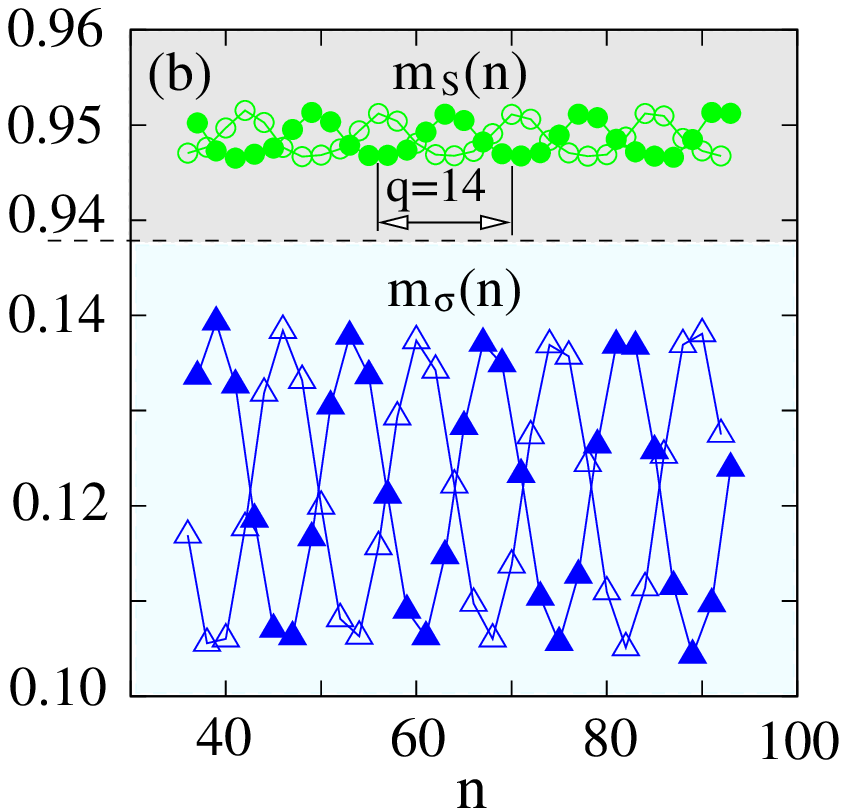}
\end{minipage}
\caption{(Color online)
(a) DMRG results for the GS entanglement entropies  \textit{vs} $l$ for
even- (open circles) and odd-cell (filled circles) subblocks $l$ at
$t=135^{\circ}$ (OBC), exhibiting oscillation periods $q=14$. The line
shows the theoretical result, Eq.~(\ref{EE}), with $c=1$.
(b) The local magnetic moments [$m_S(n)$ and $m_{\sigma}(n)$]
of the same system for even- (open circles) and odd-cell (filled circles) 
indeces $n$ exhibit the same  oscillation periods ($q=14$).      
} 
\label{Ent135}
\end{figure}
%----------------------------------------
\begin{figure}
\begin{minipage}[b]{0.45\linewidth}
\centering
\includegraphics*[clip,width=41mm]{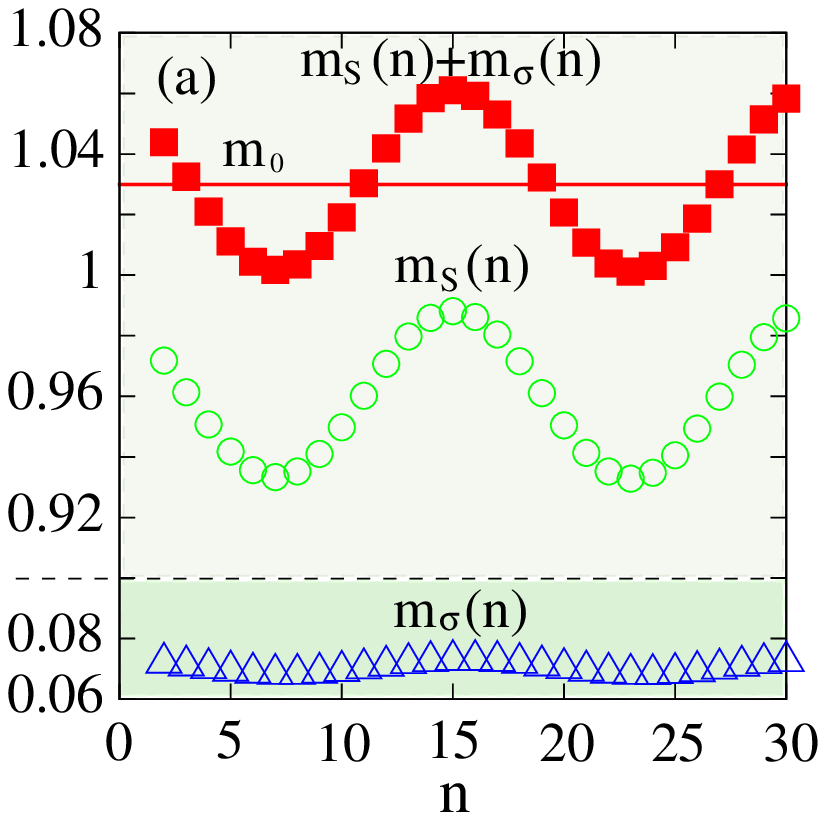}
\end{minipage}
\hspace{0.2cm}
\begin{minipage}[b]{0.45\linewidth}
\centering
\includegraphics*[clip,width=40mm]{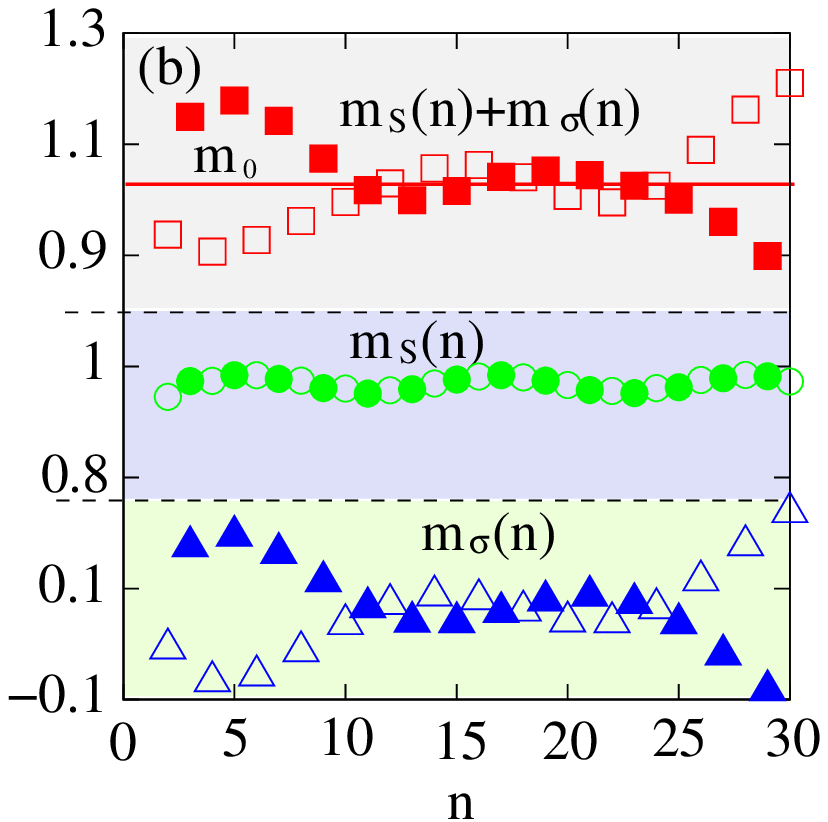}
\end{minipage}
\caption{(Color online) DMRG results for the local magnetizations
$m_S(n)=\langle S_{2n}^z\rangle$, $m_{\sigma}(n)=\langle \sigma_{2n+1}^z\rangle$
and $m_0(n)=m_S(n)+m_{\sigma}(n)$ \textit{vs} cell index $n$ for (a) periodic
and (b) open $(1,\frac{1}{2})$ chains ($t=123^{\circ}$, $L=32$, $M=33$).  
$m_0=M/L$ is the GS magnetization.
The open/filled symbols in (b) correspond to even/odd $n$. The magnetization
profiles in (b) suggest a R-type  OBC (i.e., $\sigma$ spin on the right
end). Note the  different scales of $y$ axes of both plots.       
} 
\label{M123}
\end{figure}
%--------------------------------------------------------

Another intriguing feature of the entanglement entropies 
under OBC shown in Figs.~\ref{Ent135_146} and \ref{Ent135}(a) 
is their  periodic structure. At $t=146^{\circ}$, the period
$q=8$ of $S(L,l)$ coincides with the expected period  of the 
plateau state  with magnetization $m_0=\frac{9}{8}$ 
[see Eq.~(\ref{rule})]. The  common origin  of both periods
is further  supported by the  numerical results  for the   
entropy $S(L,l)$ and  corresponding magnetization profiles 
of an open chain at $t=135^{\circ}$, Fig.~\ref{Ent135}(a). 
Using these results, it may be  speculated that the actual 
GS  at $t=135^{\circ}$ is the plateau $m_0=\frac{15}{14}$ 
state, since the other possible 
values of $m_0$, admitted by Eq.~(\ref{rule}), deviate 
significantly  from the DMRG estimate $m_0\approx 1.07$.
The periodic  magnetic patterns in Fig.~\ref{Ent135}(b), corresponding to even and odd
elementary cells,  are shifted by a half period. Another obvious effect of the
boundaries is the enhancement of the amplitudes of oscillations, especially
those related to the magnetic moment  $m_{\sigma}(n)$. 
Actually, the  absence of  visible  periodic structures in the  DMRG data for $S(L,l)$ 
in periodic chains, Fig.~\ref{Ent135_146},  is probably due to  the extreme smallness of  
the oscillation amplitudes under PBC.    

On approaching  the transition point $t_2$, the boundary effects in open chains become
stronger. A comparison between the established magnetic structures in 
periodic and open chains at $t=123^{\circ}$ is presented in  Fig.~\ref{M123}. 
An  important observation in the case of  periodic chains,
Fig.~\ref{M123}(a),  is the strongly enhanced amplitude of the $m_S(n)$ oscillations,
which dominates  by an order of magnitude the amplitude related to the
$\sigma$ spins.  
Note that the amplitude and the profile  of $m_S(n)$ remain almost
unchanged in both cases, apart from a  phase shift and some modifications 
close to the end spins. On the contrary, as seen in Fig.~\ref{M123}(b), 
the OBC  notably modifies the magnetic structure related with the  
$\sigma$ spins, the most impressive being the strong enhancement of the
local magnetizations $m_{\sigma}(n)$ 
spreading deep in the bulk.

In conclusion, we find a convincing numerical support for a
plateau $m_0=\frac{9}{8}$ state close to the FM phase boundary
$t=t_F$. Due to strong FSS effects, it  is 
difficult to track the development of this magnetic 
state as $t$ is changed down to $t=t_2$, where the 
phase transition  to  a non-magnetic state takes place.  
As in the case of $t_1^{'}<t<t_1$, we may speculated that
the region is occupied by an incommensurate 
Luttinger-liquid magnetic state. However, as seen from 
the DMRG data at $t=135^{\circ}$, the scenario with some 
intermediate plateau states can not be definitely excluded.  
%------------------------------------------
\begin{figure}
\centering
\includegraphics*[clip,width=75mm]{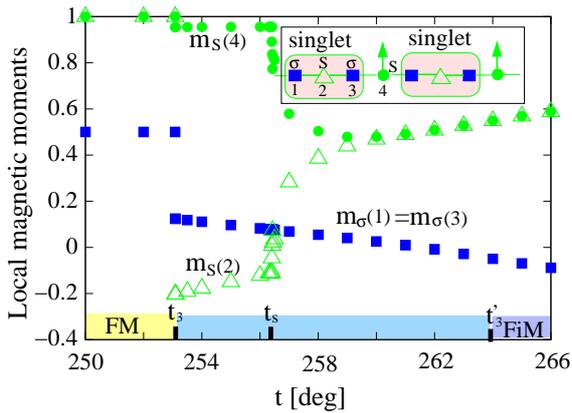}
\caption{(Color online) 
DMRG results ($L=32, PBC$) for the local magnetic moments \textit{vs} $t$ 
of the doubly degenerate FiM state appearing between the FM and FiM phases. 
The Inset shows a cartoon of the state in the 
region $t_3<t<t_s$. $t_3=253.08^{\circ}$, $t_3^{'}= 264.0^{\circ}$, 
and $t_s=256.4^{\circ}$.
} 
\label{M250-275}
\end{figure}
%------------------------

%%%%%%%%%%%%%%%%%%%%%%%%%%%%%%%%%%%%%%%%%%%%%%%%%%%%%%%%%%%%%%%%%%%%%%%%%%
(iii) \textit{Degenerate FiM phase in the sector $t_3<T<t_3^{'}$}:

Apart from the   shift $\frac{3\pi}{2} \rightarrow t_3$ of the FM boundary (see
Fig.~\ref{diagram}), quantum fluctuations also  stabilize a 
new  doubly degenerate FiM  phase in
the  vicinity of $t_3$  ( $t_3<t<t_3^{'}$). 
Here $t_3^{'}=  264.0^{\circ}$ is the new boundary of the  FiM phase.
 In Fig.~\ref{M250-275}, we show DMRG results for the local magnetic moments
$m_S(n)$ and $m_{\sigma}(n)$ in two neighboring cells ($n=1,2$). Unlike 
the standard  FiM phase, where the magnetic moment is uniformly  distributed
between the lattice cells [i.e., $m_0(n)=\frac{1}{2}$], the period of the discussed  degenerate 
FiM state includes  two lattice  cells, where $m_0(n)+m_0(n+1)=1$,
$m_{\sigma}(n)=m_{\sigma}(n+1)$, and $m_S(n)\neq m_S(n+1)$.  
  The transition  to the FM phase at $t_3$,  a result of level crossing, 
takes place through  abrupt changes of the local magnetic moments. 
On the contrary, the transition to the FiM phase at $t_3^{'}$ is preceded by a smooth decrease
to zero of the order parameter $\delta m=|m_S(n+1)-m_S(n)|$,  $\delta
m(t_3^{'})=0$. The gap $\Delta (t)$ between the FiM GS and the excited degenerate
state vanishes at the critical point $t_3^{'}$. At $t=t_3^{'}$, the
gap scales to zero as  $\Delta_L(t_3^{'})\propto 1/L$. 

The special point  $t_s$ (where the sign  of the correlator  
$\langle \bs{S}_{2n}\cdot\bs{\sigma}_{2n\pm 1}\rangle$ is
changed) divides the interval  $t_3<t<t_3^{'}$ into two regions
with different behaviors of  the local magnetic moments. Although degenerate, 
for $t>t_s$ the magnetic structure  resembles the classical N\'{e}el  phase.
In the Inset of Fig.~\ref{M250-275}, we show a cartoon of the state in the region
$t_3<t<t_s$ implied by an analysis of the SR correlations close to $t_s$. 
Due to the extreme weakness of the nearest-neighbor 
correlators $\langle \bs{S}_{2n}\cdot\bs{\sigma}_{2n\pm 1}\rangle$
($n=2,4,\ldots$), one half of the $S$ spins forms  an almost saturated magnetic
state. On the other hand, the rest of the spins is divided into  three-spin
clusters,  which exhibit SR correlations typical for the cluster singlet 
state in Table~\ref{t}.

%%%%%%%%%%%%%%%%%%%%%%%%%%%%%%%%%%%%%%%%%%%%%%%%   
\subsection{The critical spin-liquid phase (SL)} 
%%%%%%%%%%%%%%%%%%%%%%%%%%%%%%%%%%%%%%%%%%%%%%%%%%%%5

%figure=====================================
\begin{figure}
\begin{minipage}[b]{0.45\linewidth}
\centering
\includegraphics*[clip,width=40mm]{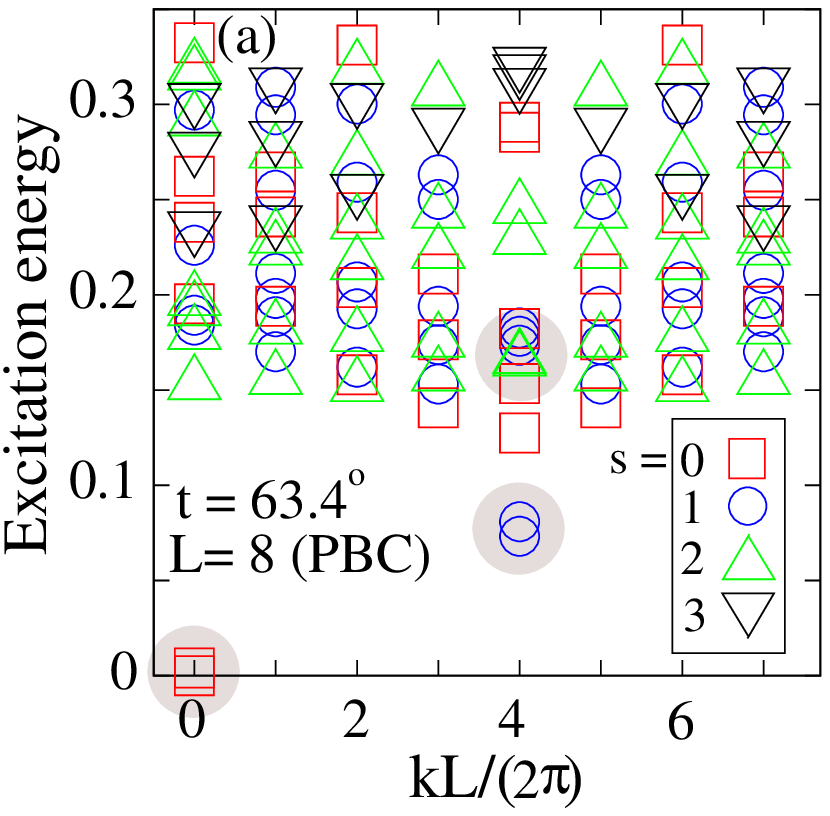}
\end{minipage}
\hspace{0.2cm}
\begin{minipage}[b]{0.45\linewidth}
\centering
\includegraphics*[clip,width=42mm]{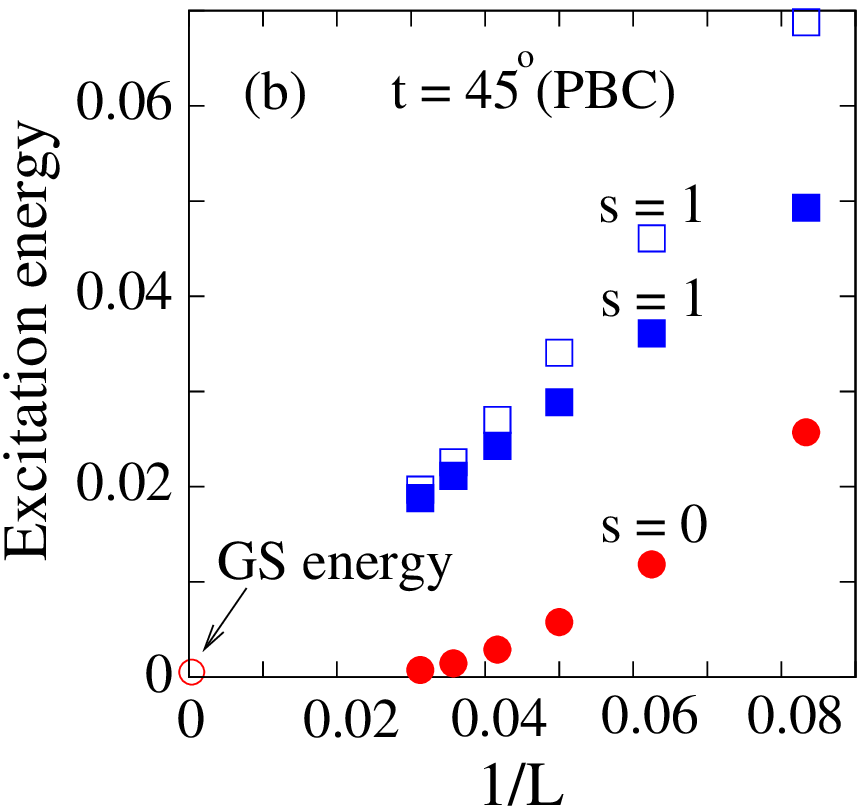}
\end{minipage}
\caption{(Color online) 
(a) Numerical ED results for the low-lying (spin $s=0,1,2$, and
$3$) excitation energies of the  periodic $L=8$ chain at $t=63.4^{\circ}$.
The shaded symbols correspond to  the lowest pairs of singlet, triplet and
quintet excitations.
(b) DMRG results for the finite-size scaling of the lowest singlet and triplet modes at
$t=45^{\circ}$.
} 
\label{doubling}
\end{figure}
%-----------------------

As noticed above, the $uv$ clustering of the GS in the SL region in
Fig.~\ref{diagram} -- a special effect of the three-body interaction --
presumably results in a double degeneracy of the singlet GS in the large-$L$ limit.
As indicated in Fig.~\ref{doubling}(a), some low-lying larger-spin  states in the ring 
spectrum also exhibit a tendency towards formation of  quasi-degenerate pairs. 
In fact, even in the region around $t=45^{\circ}$, where the singlet gap 
in  the $L=8$ ring is relatively large, the performed FSS 
analysis  for  larger-$L$ rings supports the suggested picture  and,  in particular,  
implies an exponentially fast (with $L$) doubling of the lowest singlet and triplet states [see
Fig.~\ref{doubling}(b)]. Due to strong boundary effects, the  discussed doubling
in the lowest part of the spectrum remains invisible in open chains
up to the largest ($L=256$) simulated system.

Assuming conformal invariance, additional  properties   of  the non-magnetic SL phase 
can be extracted from the  FSS  behavior of the GS and  the lowest excited
states. Since the numerical simulations for periodic chains 
are hampered by the quasi-degeneracy of the  GS,  the following analysis is  performed under OBC. 
The expected large-$L$ behavior of the GS energy reads \cite{bloete}       
\be
\frac{E_0(L)}{L}=e_{\infty}+\frac{f_s}{L}-\frac{v_s\pi
c}{6\gamma L^2}+o(L^{-2}),
\label{FSS_e0}
\ee
where $e_{\infty}$ is the bulk GS energy per unit cell, $f_s$ is the surface free
energy ($f_s=0$ under PBC), $v_s$ is the spin velocity, $c$ is the central
charge,  and $\gamma =1,4$ for PBC and OBC, respectively.  
For OBC, the expected tower of excited states related to some primary operator is 
defined by \cite{cardy}
\be
\Delta_n(L)\equiv E_{n}(L)-E_0(L)=\frac{\pi v_s}{L}\left( x_s +n\right)
+o(L^{-1}),
\label{FSS_en}
\ee  
where $n=0,1,2,\ldots$ and  $x_s$ is the universal surface exponent related to the
same operator. The exponent $x_s$ is known, in particular, for the energy states of  
the isotropic spin-$\frac{1}{2}$ Heisenberg chain in  the $m$ sectors
($x_s^{(m)}=m^2$, $m=1,2,\ldots$),  where $m$ is the $z$ component of the total
spin \cite{alcaraz}.     The Hamiltonian (\ref{h}) respects the spin-rotation symmetry, 
so that the above asymptotic expression, when used as a fitting  ansatz,  
have to be supplemented by appropriate logarithmic terms \cite{hallberg}.
%=================================
\begin{figure}
\begin{minipage}[b]{0.45\linewidth}
\centering
\includegraphics*[clip,width=40mm]{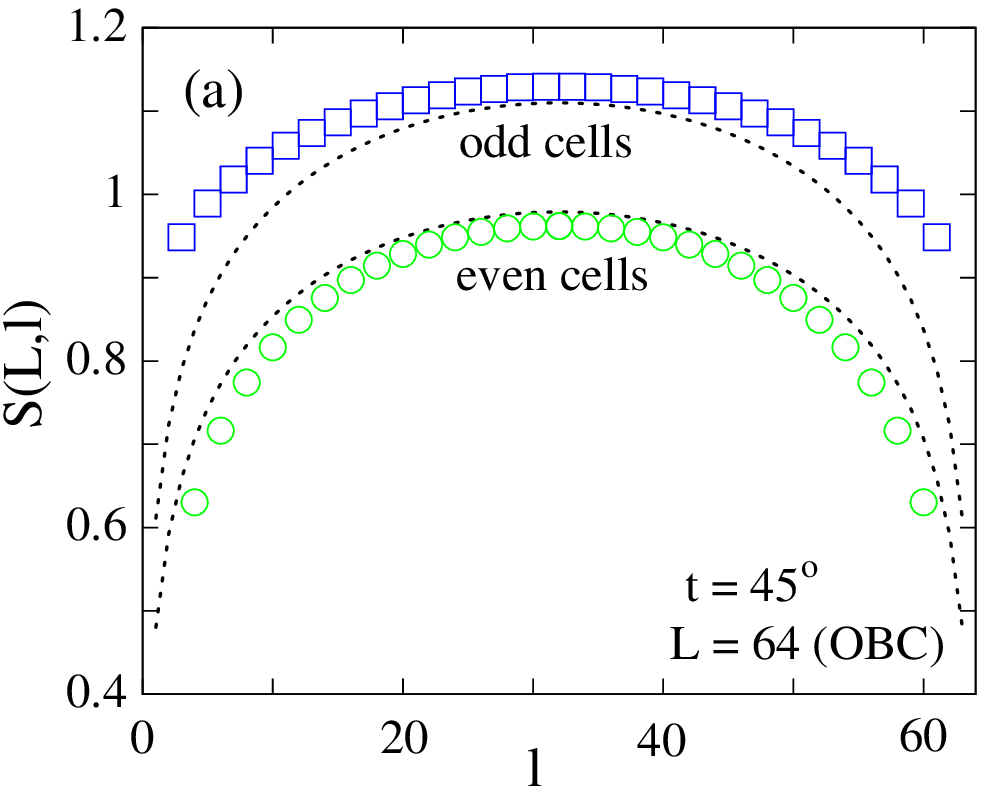}
\end{minipage}
\hspace{0.2cm}
\begin{minipage}[b]{0.45\linewidth}
\centering
\includegraphics*[clip,width=30mm]{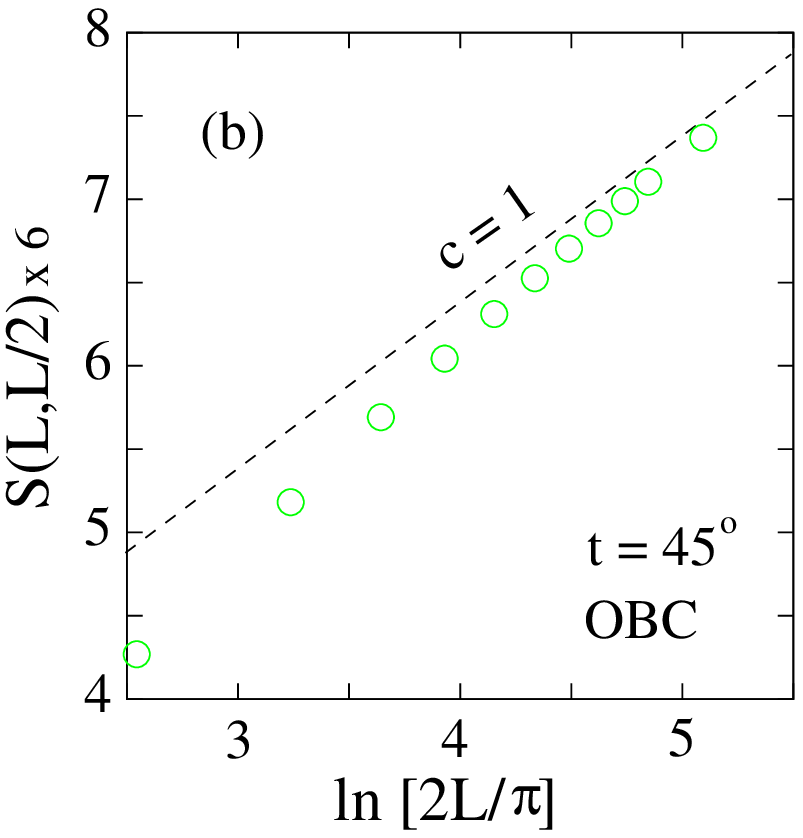}
\end{minipage}
\caption{(Color online)
(a) Numerical DMRG results for the even- and odd-cells entanglement entropy of  
the SL state ($t=45^{\circ}$) \text{vs} the number of subblock cells $l$.
The dot lines represent the theoretical result in Eq.~(\ref{EE}) with $c=1$.
(b) Extrapolation of the numerical results for the  $l=L/2$ 
entanglement entropy $S(L,L/2)$ \text{vs} $\ln (2L/\pi)$ at $t=45^{\circ}$. 
The dot line corresponds to $S(L,L/2)$ with the central charge $c=1$.
} 
\label{Ent45}
\end{figure}
%=================================
Since the central charge $c$ can be independently obtained from a fit 
of Eq.~(\ref{EE}) to the DMRG data, the asymptotic expression for 
$E_0(L)$ can, in principle,  be used to find the non-universal parameters  
$e_{\infty}$, $f_s$ and $v_s$. Thus, the surface exponents $x_s$ of
different primary operators can be extracted by fitting Eq.~(\ref{FSS_en}) to
the numerical data. However, due to logarithmic corrections, the precise estimate 
of $v_s$ from Eq.~(\ref{FSS_e0}) for isotropic systems may 
require numerical simulations  of extremely large systems. 

In Fig.~\ref{Ent45}(a), we show DMRG results for the GS entanglement entropy $S(L,l)$  
of the open alternating-spin chain at $t=45^{\circ}$ ($L=64$). We observe two
different branches of $S(L,l)$ corresponding to even and odd subblock lengths
$l$. Similar even-odd  oscillations in the entanglement entropy have been
firstly reported in  open spin-$\frac{1}{2}$ XXZ chains,  
including the isotropic limit \cite{lafforencie}.
In this work, it was  clarified  that the alternating part of $S(L,l)$,  
decaying  away from the boundary with a
universal power law, appears as a result of oscillations of the energy density. 
Further, the  latter  oscillations were related with the tendency 
of the critical system towards formation of local singlet bonds, combined with the 
strong tendency of the  end spins to form local singlets.     
As the extrapolation of the numerical data for $S(L,L/2)$ \textit{vs} $\ln
\left( 2L/\pi\right)$ up to $L=256$ suggests a
critical behavior  with central charge $c=1$ [see Fig.~\ref{Ent45}(b)],
to understand the even-odd effect in the  alternating spin 
chain one may suggest a scenario similar. 
However, the  picture looks more complex as the formation of local singlet
states in the alternating spin model  includes at least four  neighboring
spins.
%figure=====================================
\begin{figure}
\begin{minipage}[b]{0.45\linewidth}
\centering
\includegraphics*[clip,width=40mm]{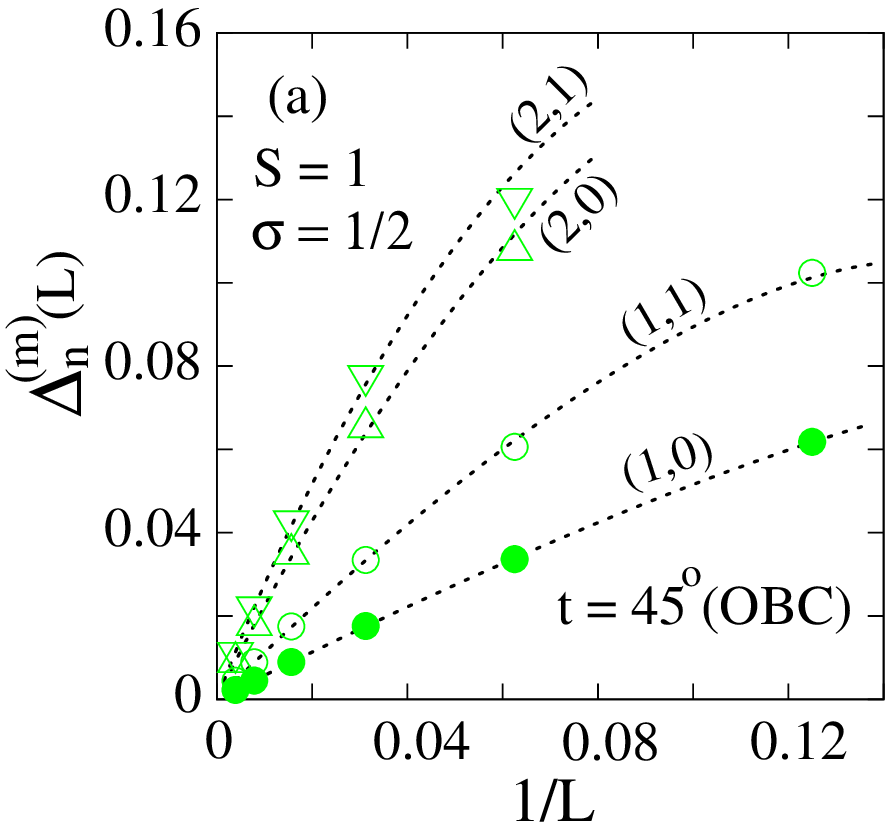}
\end{minipage}
\hspace{0.2cm}
\begin{minipage}[b]{0.45\linewidth}
\centering
\includegraphics*[clip,width=40mm]{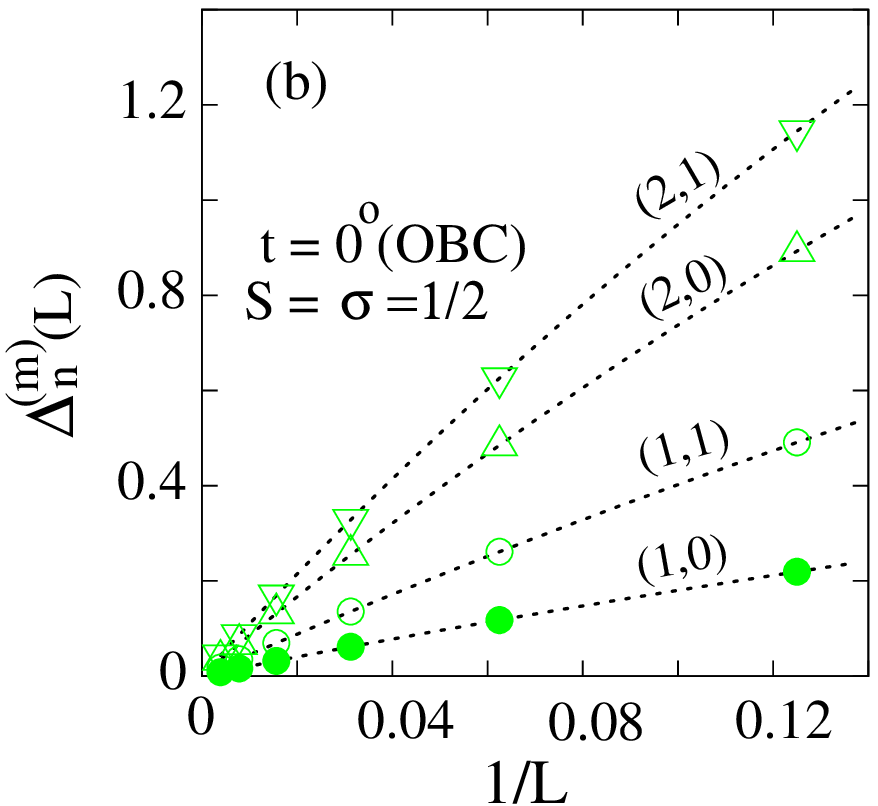}
\end{minipage}
\caption{(Color online)
(a) Scaling of the lowest triplet [$(m,n)=(1,0)$ and $(1,1)$] and quintet 
[$(m,n)=(2,0)$ and $(2,1)$] excitation gaps in (a) the alternating spin $S=1$
and $\sigma=\frac{1}{2})$ open chain (DMRG, OBC, $t=45^{\circ}$). (b) Scaling 
of the related gaps in the uniform spin ($S=\sigma =\frac{1}{2}$)
model without three-body terms (DMRG, OBC, $t=0^{\circ}$). 
The dashed  lines show the best fits  of  $\Delta_n^{(m)}(L)=E_n^{(m)}(L)-E_0(L)$
to the DMRG data (symbols) obtained by Eq.~(\ref{ansatz}. 
} 
\label{scaling}
\end{figure}
%==================

In Fig.~\ref{scaling}, we compare FSS results for the lowest two excitations  in
the triplet ($m=1$) and quintet ($m=2$) towers of states of the Hamiltonian ${\cal
H}_{\sigma S}$, Eq.~(\ref{h}), for two cases:  
(i)  $(S,\sigma)=(1,\frac{1}{2})$ at $t=45^{\circ}$ 
and (ii) $(S,\sigma)=(\frac{1}{2},\frac{1}{2})$ at $t=0^{\circ}$.   
The fit of the reduced gaps $L\Delta_n^{(m)}$  is performed  by the four-parameter ansatz
\be\label{ansatz}
L\Delta_n^{(m)}(L)=a_n^{(m)}+\frac{b_n^{(m)}}{\ln\left(L/\xi_{n}^{(m)}\right)}+\frac{c_n^{(m)}}{L}.
\ee
For systems belonging to the Gaussian universality class -- like the
isotropic spin-$\frac{1}{2}$ chain in the second case -- the first fitting parameter 
$a_n^{(m)}$ is expected to approach  the exact result $a_n^{(m)}=\pi
v_s(m^2+n)/2$, which gives $[m,n]\equiv a_n^{(m)}/a_0^{(1)}=m^2+n$.
In fact, the performed fits for the $(S,\sigma)=(\frac{1}{2},\frac{1}{2})$ 
chain give  the numerical estimates  [1,1]=1.99, [2,0]=3.99, and
[2,1]=4.96, which  excellently reproduce the expected theoretical ratios. 
Moreover, a comparison of Eqs.~(\ref{FSS_en}) and (\ref{ansatz}) 
implies the relation $a_0^{(1)}=v_s/2$, which gives  an estimate 
for $v_s$ deviating only by about $0.6\%$ from the exact result $\pi/2$. 
For the alternating-spin chain at $t=45^{\circ}$, 
similar fits  give the numerical estimates 
$v_s=0.38$, [1,1]=2.11, [2,0]=4.44, and [2,1]=5.69.  
In spite of the larger deviations from the theoretical results for $[m,n]$, 
the  observed  structure of the lowest-lying part of the spectrum  in the
alternating-spin model remains close to the structure  in the reference 
spin-$\frac{1}{2}$ Heisenberg chain. As may be expected,
in the middle of the range occupied by the SL phase, where
the doubling (with L) of the lowest singlet and triplet states 
is faster (see Fig.~\ref{gaps}), the deviations of $[m,n]$ 
from the expected theoretical results are smaller. For example, at 
$t=63.4^{\circ}$ the same fitting procedure gives $v_s=1.88$,
$[1,1]=1.93$, $[2,0]=4.00$, and $[2,1]=5.35$. 

The established one-to-one mapping of the lowest-lying excitations of both
models suggests similar critical properties. Since the unit cell of 
the reference spin-$\frac{1}{2}$ model  contains two equivalent lattice sites,
under PBC this means a doubling of the spectrum and, in particular,
two equivalent critical modes. This explains the discussed  
doubling of the lowest-lying excitations in the alternating-spin 
$(1,\frac{1}{2})$ ring. Thus, both  the GS entanglement 
entropy as well as  the FSS properties of the SL phase  
point towards a Gaussian type critical behavior.
What is changed in the region occupied
by the SL phase is the non-universal parameter $v_s$.
Since the alternating-spin $(1,\frac{1}{2})$ ring  exhibits two  
equivalent critical modes, the SL state  may be interpreted  as a critical 
spin-liquid phase  described by two  Gaussian conformal theories associated 
with these modes. Similar critical phases  have been studied in some  
exactly solvable models, including spin models with extra 
three-body exchange interactions.  In particular, there is an 
exactly solvable alternating-spin $(S,\sigma)$ model \cite{aladim} 
closely related to the  generic model discussed in this work at the point 
$t=45^{\circ}$. In fact, the difference  between both models  is  reduced to
an additional  FM exchange term
($h_n^{\sigma\sigma}=J_3\bs{\sigma}_{2n-1}\cdot\bs{\sigma}_{2n+1}$,
$J_3<0$) in the exactly solvable model.  Assuming that $h_n^{\sigma\sigma}$ represents  
an irrelevant operator (in a renormalization group  sense), it may be speculated 
that both models exhibit similar
critical properties. In particular, for  the exactly solvable
$(S,\sigma)$ model,  it has been predicted \cite{aladim} that the critical behavior 
can be described by an  effective central charge which is the sum of the central
charges related with two critical modes, t.e., $c_{eff}=3\sigma /(\sigma
+1)+3(S-\sigma)/(S-\sigma+1)$. In the special case $(S,\sigma)=(1,\frac{1}{2})$ this 
gives  $c_{eff}=1+1=2$, which coincides  with the expected  critical
behavior of the SL phase.

\subsection{Critical nematic-like phase (N)}

The behavior of both the low-lying excitations and SR correlations 
indicate the existence of a different  non-magnetic 
phase  in the parameter region $\frac{\pi}{2}\simeq t\simeq \frac{2\pi}{3}$ (N sector in
Fig.~\ref{diagram}). Indeed, as seen in Fig.~\ref{gaps}, in the
vicinity of $t\simeq \frac{\pi}{2}$ the  quintet ($s=2$, $k=0$) excitation is strongly
softened  
 and becomes the lowest excited state up to $t\simeq \frac{2\pi}{3}$. 
Moreover, the DMRG calculations for
somewhat larger periodic systems (up to $L=28$) reveal the same  structure of
the low-lying part of the spectrum. Unfortunately, slow convergence of the  DMRG
method in this region hampers   more extensive numerical simulations of the
FSS  properties of the excitation gaps. 
The picture of SR correlations in this region, Fig.~\ref{src},  allows to speculate that the
properties of the non-magnetic state  are
mainly controlled by the $S$ subsystem. Indeed, as already discussed above, the correlator
 $\langle\bs{S}_{2n}\cdot\bs{S}_{2n+2}\rangle$ exhibits a strong modification
 in a vicinity of the second phase boundary ($t=t_2$). Meanwhile, the behavior of the SR correlator  
$\langle\bs{\sigma}_{2n-1}\cdot\bs{\sigma}_{2n+1}\rangle$ 
remains practically unchanged in the entire N sector, including
the  regions around both phase boundaries. Interestingly, in the entire
N sector the typical values  of the latter correlator remain  relatively close to the value
$\frac{1}{4}-\ln (2)\approx -0.443$ characteristic for the isotropic
spin-$\frac{1}{2}$ chain. Another peculiarity in this region is the  extremely
week correlation between the nearest-neighbor $S$ and $\sigma$ spins [see
Fig.~\ref{src}(b)].       
%=================================
\begin{figure}
\begin{minipage}[b]{0.45\linewidth}
\centering
\includegraphics*[clip,height=45mm]{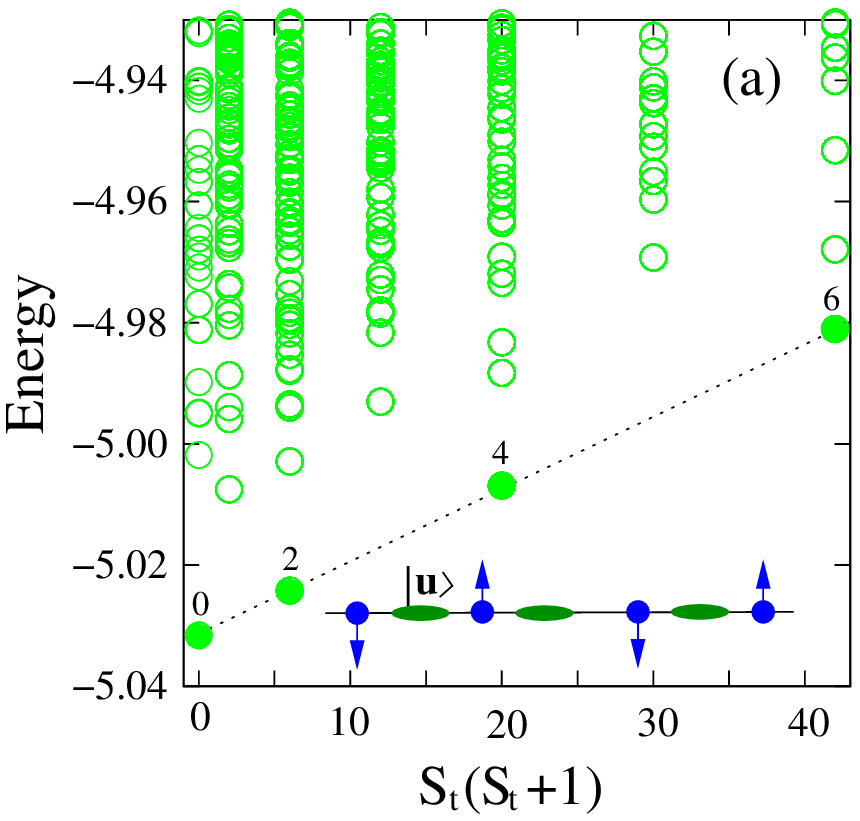}
\end{minipage}
\hspace{0.2cm}
\begin{minipage}[b]{0.45\linewidth}
\centering
\includegraphics*[clip,height=45mm]{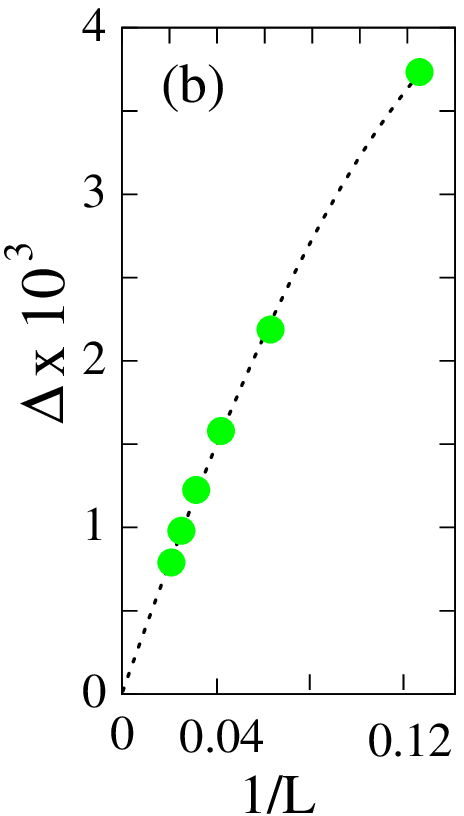}
\end{minipage}
\caption{
(Color online)
(a) Numerical ED results for the energy spectrum of the $(1,\frac{1}{2})$ model,
Eq.~(\ref{h}), at $t=110^{\circ}$ 
as a function of $S_t(S_t+1)$ ($L=8$, PBC). The numbers
denote the spins ($S_t$) of the lowest-lying multiplets (filled circles) in the even $S_t$
sectors. The dashed line is a guide for the eye. Inset: 
Cartoon of the suggested nematic-like state in the N sector. The ellipses
denote local nematic states on the even $S$ sites.
(b) Finite-size scaling of the lowest $S_t=2$ gap. The dashed line denotes
the  least-squares fit to the DMRG data ($t=110^{\circ}$, OBC) 
obtained by the fitting ansatz $L
\Delta(L)=a_0+a_1/L+a_2/L^2$.   
} 
\label{tower}
\end{figure}
%=================================
  
Further information about  the  non-magnetic (N) state can be extracted form 
Fig.~\ref{tower} showing  ED results for the excitation spectrum 
of the same system at $t=110^{\circ}$ in different total-spin ($S_t$) sectors. 
An obvious  feature of the presented  spectrum is the   
established tower of well-separated  lowest  multiplets containing only even $S_t$
sectors. Furthermore, the energies in the tower scale as  $E(S)\propto S_t(S_t+1)$. 
The observed structure is known as a fingerprint of  the spin quadrupolar
(i.e., nematic) order \cite{lauchli2}, unlike  the Anderson
tower -- a characteristic 
of the  N\'{e}el  order -- containing all $S_t$ sectors \cite{anderson}.
In fact, Anderson towers of states have been observed  even in  some  
finite  isotropic spin-$S$  chains and  magnetic molecules \cite{schnack}, 
including  spin-$\frac{1}{2}$ Heisenberg rings which support
a quasi-long-range  N\'{e}el-type  order in the 
thermodynamic limit \cite{eggert}. In the  same spirit,  
we consider  the specific structure in  Fig.~\ref{tower}(a) 
as  a fingerprint of  a non-magnetic state with dominant quadrupolar spin fluctuations. 
The  FSS  of the quintet excitation gap $\Delta (L)\propto 1/L$, Fig.~\ref{tower}(b), 
pointing  towards a gapless N state, is consistent with the above suggestion. 

The Inset of Fig.~\ref{tower}(a) shows the  cartoon  of a 
tentative nematic-like  state  respecting  the established properties 
of the low-lying spectrum and 
the peculiarities of the SR correlations. In the vector basis
$|\alpha\rangle$ ($\alpha =x,y,z$)   of the  spin-1
operators $\bs{S}_{2n}$, 
an arbitrary on-site quadrupolar state can be written in the form 
$|\bs{u}_{2n}\rangle=\sum_{\alpha}u_{2n}^{\alpha}|\alpha\rangle$, where $\bs{u}_{2n}$ is a real 
unit vector.  Since $\langle \bs{u}_{2n}|S_{2n}^{\alpha}|\bs{u}_{2n}\rangle =0$
for every $\alpha$, the $|\bs{u}_{2n}\rangle$ states on the even sites
vanish the nearest-neighbor $\sigma S$ correlations for an arbitrary
configuration of  the $\sigma$ spins, in accord with the established extremely 
weak nearest-neighbor  $\sigma S$ correlations. To reveal the origin of the
observed strong AFM nearest-neighbor $\sigma\sigma$ correlations,  it is
instructive to recast the local three-body exchange term in Eq.~(\ref{h}), which  
dominates the interactions in the N region, to the following symmetric form
\be\label{h3}
h_n^{(3)}=J_2 \bs{\sigma}_{2n-1}\cdot \hat{\bs{Q}}_{2n}\cdot
\bs{\sigma}_{2n+1},\, J_2>0,
\ee
where the  symmetric tensor $Q_{2n}^{\alpha\beta}=S_{2n}^{\alpha}S_{2n}^{\beta}
+S_{2n}^{\beta}S_{2n}^{\alpha}$  is closely related to the on-site
quadrupolar order-parameter operator for the $\bs{S}_{2n}$ spins (see, e.g.,
Ref.~\cite{frustration}). Using the relation  
$\langle\bs{u}_{2n}|Q_{2n}^{\alpha\beta}|\bs{u}_{2n}\rangle
=\delta^{\alpha\beta}-u_{2n}^{\alpha}u_{2n}^{\beta}$, the effective zeroth-order 
Hamiltonian for the $\sigma$ subsystem reads
\linebreak
$\bar{h}_n^{(3)}\equiv\langle \bs{u}_{2n}|h_n^{(3)}|\bs{u}_{2n}\rangle=
J_2 \bs{\sigma}_{2n-1}^{\perp}\cdot \bs{\sigma}_{2n+1}^{\perp}$,
where $\bs{\sigma}_{2n-1}^{\perp}$ and $\bs{\sigma}_{2n+1}^{\perp}$ are the
transfer components of the $\sigma$ spins in respect to the local vector
$\bs{u}_{2n}$. $\bar{h}_n^{(3)}$ defines a kind of AFM spin-$\frac{1}{2}$ XX
model with the local quantization axis $\bs{u}_{2n}$.

In conclusion, both the SR correlations as well as  
specific structure of the low-lying excitations
point towards the establishment of an intriguing critical nematic-like phase in the
N region of the phase diagram, Fig.~\ref{diagram}, which is characterized by
quadrupolar $S$-spin fluctuations. The three-body interaction 
plays a dominant role, whereas the role of the 
FM bilinear terms ($J_1<0$) is to reduce the AFM correlations 
between the $S$ and $\sigma$ subsystems. Further properties of
this phase as well as more precise estimates for the phase boundaries
require other methods (e.g., larger-scale numerical ED simulations), 
which are beyond the scope of the present work. 
%%%%%%%%%%%%%%%%%%%%%%%%%%%%%%%%%%%%%%%%%%%%%%%%%%%%%%%%%%%
\section{Summary}
We have established the general structure of the quantum phase diagram of a
generic 1D isotropic spin model with competing biquadratic three-body exchange
interactions, with an  emphasis on the minimal model with alternating $S=1$
and $\sigma=\frac{1}{2}$ spins.  A number of observed effects as well as specific phases
(like the  doubly degenerate FiM state, the two-critical-modes  spin-liquid, as well as the
nematic-like phase) can be attributed to  peculiarities of the  three-body
exchange interaction,  such as the promotion of collinear spin configurations 
and pronounced tendency towards a nearest-neighbor clustering of the spins.
It may be expected that most of the predicted effects and  phases 
persist (or are stabilized) in higher space dimensions.      
On the experimental side, we believe that the presented results will
encourage the search for real systems  exhibiting three-body exchange
interactions. In this respect, the alternating-spin materials with complex
unit cells  constitute a  promising  background of systems 
principally allowing  manipulations of the higher-order exchange interactions.  

%%%%%%%%%%%%%%%%%%%%%%%%%%%%%%%%%%%%%%%%%%%%%%%%%%%%%%%%%%%%%%%%%%%%%%%%
\section*{Acknowledgment}

This work was supported by the Deutsche Forschungsgemeinschaft
(436BUL 113/106/0 \& SCHN 615/20-1) and the Bulgarian
Science Foundation (Grant No. F817/98).
%%%%%%%%%%%%%%%%%%%%%%%%%%%%%%%%%%%%%%%%%%%%%%%%%%%%%%%%%%%%%

%%%%%%%%%%%%%%%%%%%%%%%%%%%%%%%%%%%%%%%%%%%%%%%%%%%%%%%%%%%
\end{document}